\documentclass[traditabstract,a4paper,referee]{aa}
\usepackage{graphicx}
\usepackage{url}
  
\usepackage{times}
\newcommand{\MSun}{\mbox{M$_\odot$}}
\newcommand{\LSun}{\mbox{L$_\odot$}}

\def\apgt{\ {\raise-.5ex\hbox{$\buildrel>\over\sim$}}\ }
\def\aplt{\ {\raise-.5ex\hbox{$\buildrel<\over\sim$}}\ }
\def\lteq{\ {\raise-.5ex\hbox{$\buildrel<\over-$}}\ }

\newcommand{\Al}{$^{26}$Al\,}
\newcommand{\Fe}{$^{60}$Fe\,}

\newcommand{\AU}{au}

\begin{document}

\title{The consequences of a nearby supernova on the early Solar System}

\author{
 S.\, Portegies Zwart\inst{1}
 \and
 I.\, Pelupessy\inst{2}
 \and
 A.\, van Elteren\inst{1}
 \and
 T.P.G.\, Wijnen\inst{1,3}
 \and
 M.\, Lugaro\inst{4}
}

\offprints{S. Portegies Zwart}
\mail{spz@strw.leidenuniv.nl}
\institute{
 $^1$Leiden Observatory, Leiden University, PO Box 9513, 2300
   RA, Leiden, The Netherlands\\
 $^2$ Centrum Wiskunde \& Informatica, Science Park 123, 1098 XG
   Amsterdam, the Netherlands\\
 $^3$ Department of Astrophysics/IMAPP, Radboud University
   Nijmegen, P.O. Box 9010, 6500 GL Nijmegen, the Netherlands\\
 $^4$ Konkoly Observatory, Research Centre for Astronomy and Earth
   Sciences, Hungarian Academy of Sciences, H-1121 Budapest, Hungary
}

\date{Received / Accepted }
\titlerunning{Supernova Near the Solar System}
\authorrunning{Portegies Zwart et al.}

\abstract{ If the Sun was born in a relatively compact open cluster
  (half-mass radius $\aplt 3$\,pc) with $\apgt 10^3$ stars, it is
  quite likely that a massive ($\apgt 10$\,\MSun) star was nearby when
  it exploded in a supernova. The repercussions of a supernova can be
  rather profound, and the current Solar System may still bear the
  memory of this traumatic event.  The truncation of the Kuiper belt
  and the tilt of the ecliptic plane with respect to the Sun's
  rotation axis could be such signatures.  We simulated the effect of
  a nearby supernova on the young Solar System using the Astronomical
  Multipurpose Software Environment.  Our calculations are realized in
  two subsequent steps in which we study the effect of the supernova
  irradiation on the circumstellar disk and the effect of the impact
  of the nuclear blast-wave which arrives a few decades later. We find
  that the blast wave of our adopted supernova exploding at a distance
  of $0.15$--$0.40$\,pc and at an angle of $35^\circ$--$65^\circ$ with
  respect to the angular-momentum axis of the circumsolar disk would
  induce a misalignment between the Sun's equator and its disk to
  $5^\circ.6\pm1^\circ.2$, consistent with the current value.  The
  blast of a supernova truncates the disk at a radius between $42$ and
  $55$\,\AU, which is consistent with the current edge of the Kuiper
  belt. For the most favored parameters, the irradiation by the
  supernova as well as the blast wave heat the majority of the disk to
  $\apgt 1200$\,K, which is sufficiently hot to melt chondrules in the
  circumstellar disk.  The majority of planetary system may have been
  affected by a nearby supernova, some of its repercussions, such as
  truncation and tilting of the disk, may still be visible in their
  current planetary system's topology.  The amount of material from
  the supernova blast wave that is accreted by the circumstellar disk
  is too small by several orders of magnitude to explain the current
  abundance of the short live radionuclide \Al.

}

\maketitle

\section{Introduction}

Stars are expected to form in clustered
environments~\citep{2003ARA&A..41...57L} and the Sun is probably no
exception~\citep{2009ApJ...696L..13P}.  From dynamical considerations
with the hypothesis that the Sun's circumstellar disk survived close
encounters, the proto-solar cluster was $\sim 1$\,pc in size,
contained $\sim 10^3$ stars, and ultimately became gravitationally
unbound~\citep{2009ApJ...696L..13P}.  In this dense stellar
environment several massive $\apgt 20$\,\MSun\, stars must have been
present and one star was probably close when it exploded in a
supernova \citep{2016MNRAS.462.3979L,2017MNRAS.464.4318N}.

The repercussions of a nearby supernova may be profound
\cite{2017MNRAS.469.1117C}, and it's effect has been studied using
grid-based hydrodynamical simulations by \citet{2007ApJ...662.1268O},
who injected a slab of dense gas at 400 km/s perpendicular to a
40\AU\, circumstellar disk. They found that the disk accretes $\sim
1$\% of the slab, which is insufficient to explain the abundance in
the early Solar System of \Al. This is a short-lived radionuclide
($t_{1/2}$=0.717 Myr) whose initial abundance corresponds to a
\Al/$^{27}$Al ratio of $5 \times 10^{-5}$, as inferred from excesses
of its decay product, $^{26}$Mg, in meteoritic materials such as
calcium- aluminium-rich inclusions (CAIs)
\citep{2008LPI....39.1999J}. Iron-60 ($t_{1/2}$=2.62 Myr) is another
short-lived radionuclide (SLR) of interest because it's presence in
the early Solar System could indication the occurrence of (one or
more) nearby core-collapse supernovae during formation. However, its
initial abundance is highly uncertain, with recently reported values
ranging from $^{60}$Fe/$^{56}$Fe$\sim10^{-8}$, which would not require
one or more local supernovae \citep{2012LPI....43.1703T,tang15}, up to
to $\sim 10^{-7} - 10^{-6}$ \citep{mishra14,telus18}, which would
require it. The traditional view is, therefore, that $^{26}$Al
originated from a $\sim$25 M$_{\odot}$ star that exploded as supernova
\citep{2005ASPC..341..515M,2006ApJ...652.1755L} however due to the
potential difficulty for supernovae to match the $^{60}$Fe/$^{26}$Al
ratio in the early Solar System other sources have been also
suggested, such as Wolf-Rayet stars
\citep{2009ApJ...705L.163G,gounelle12,dwarkadas17}. In the framework
of accretion of supernova ejecta by a protoplanetary disk, to solve
the problem of the too-small accretion efficiency Ouellette et
al. (2007, 2010) \nocite{2010ApJ...711..597O,2007ApJ...662.1268O}
proposed that the radionuclides $^{26}$Al and $^{60}$Fe were
completely condensed into large ($\mu$m size) dust grains, which have
a much higher injection efficiency.

We revisit this problem to study the more intricate interaction
between the supernova irradiation and its blast wave and the Sun's
circumstellar disk, in particular when both do not perfectly
align. Another reason to revisit this problem is to investigate to
what extend today's anomalously small disk 42--55\,\AU\, in comparison to
the 100--400\,\AU\, disks found in nearby star-forming regions
\citep{2005A&A...441..195V}, could be the result of the truncation by
the blast wave of a nearby supernova.  We report on our findings in two
sections, \S\,\ref{Sect:Irradiation} where we discuss the effect of
the supernova radiation on the circumstellar disk and
\S\,\ref{Sect:blastwave} where we discuss the hydrodynamics of the
blast wave hitting the disk, we discuss the results and consequences
in \S\,\ref{Sect:Discussion}.

\section{Effect of the supernova irradiation}\label{Sect:Irradiation}

For the supernova explosion distances we are interested in, the first
impact to which the nearby proto-planetary disk is exposed will be the
intense radiation emitted in the early stages of the supernova. This
radiation can ionize and heat the disk material and even disrupt the
disk. In order to assess the extent of this heating and whether the
disk survives, we start by calculating the effect of the supernova
irradiation on the disk using a combined hydrodynamics and radiative
transfer solver.

\begin{figure}
\includegraphics[width=1.0\columnwidth]{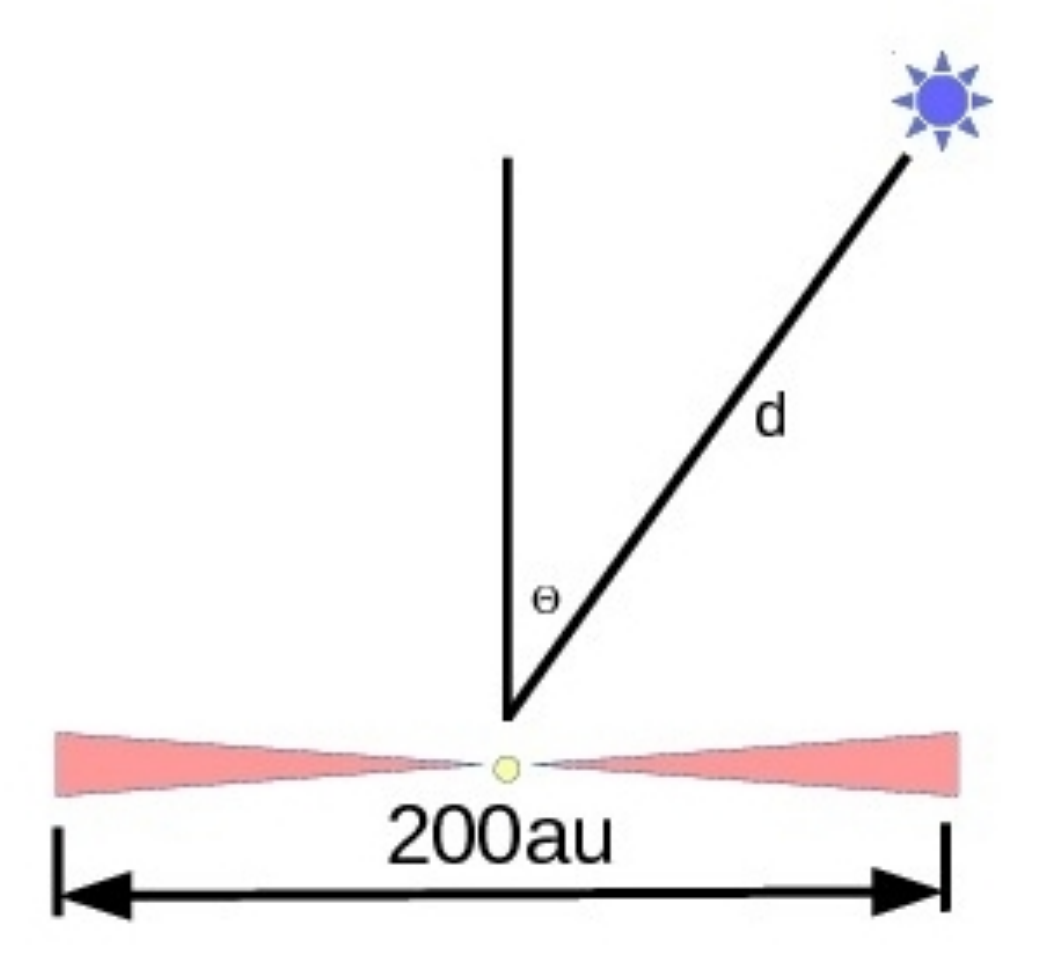}
\caption[]{Geometry of the experimental setup. The circumestellar disk
  is represented near the bottom with a diameter of 200\,au. The
  exploding star (blue star to the top right) at a distance $d$ and
  with angle $\Theta$ with respect to the disk's angular momentum
  axis.
  \label{Fig:geometry}
}
\end{figure}

\begin{figure}
\includegraphics[width=1.0\columnwidth]{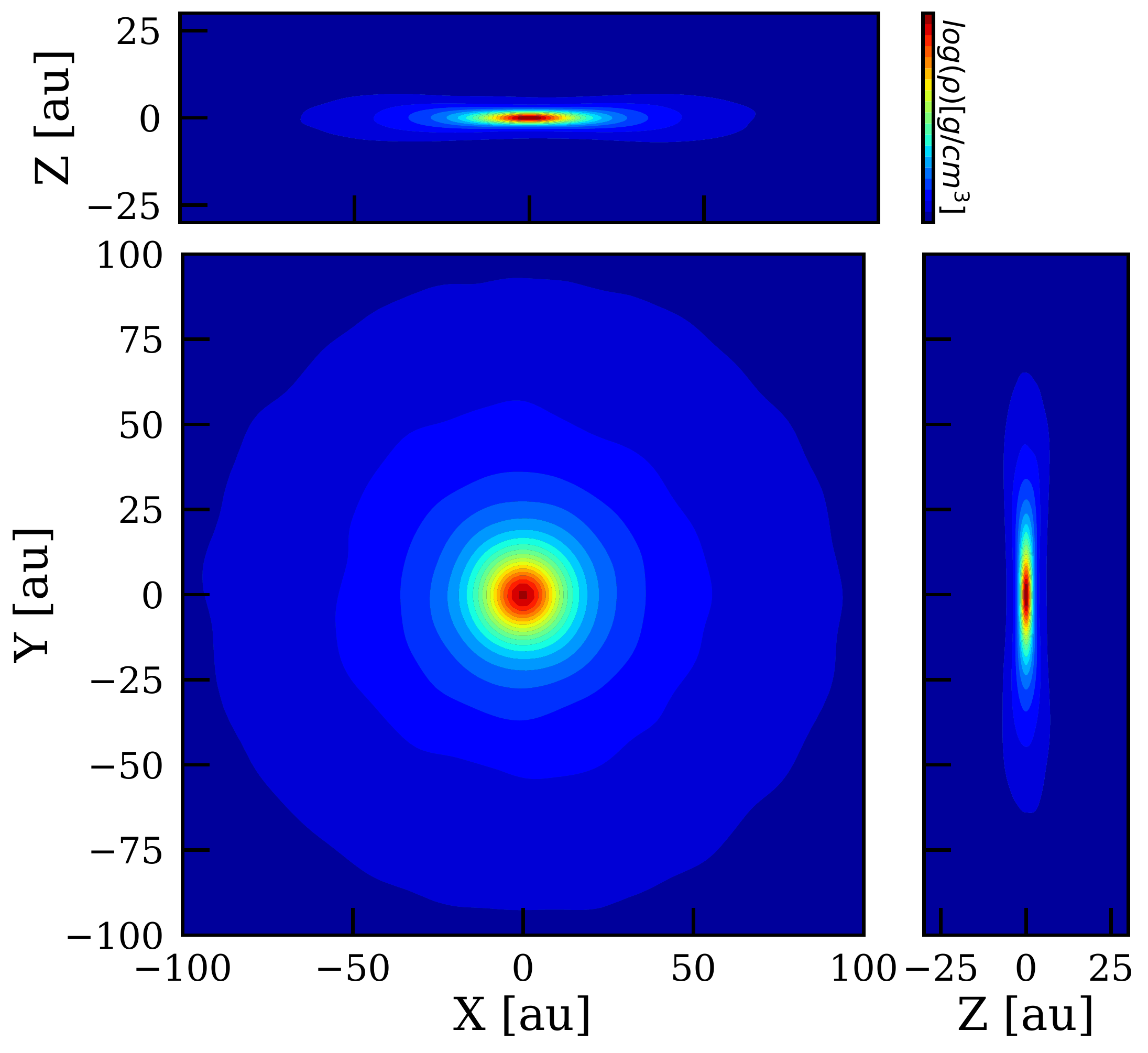}
\caption[]{Mid-plane density distribution of the protoplanetary disk
  with a top view in the middle and side views to the right and
  top. The density is color coded, the coding scheme is presented to
  the top right of the figure.
  The density extremal range from $10^{-12.9}$\, g/cm$^3$ to  $10^{-9.97}$\, g/cm$^3$. 
  \label{Fig:Disk_density}
}
\end{figure}

We start each simulation by generating the young Solar System composed
of a 1\,\MSun\, star and a 0.01\,\MSun\, gaseous disk with an inner
radius of $r_{\rm min} = 1$\,\AU\, and an outer radius of $r_{\rm max}
= 100$\,\AU.  In Fig.\,\ref{Fig:geometry} we present the geometry of
our simulation setup, and Fig.\,\ref{Fig:Disk_density} shows three
projections of the initial disk.

\begin{figure}
\includegraphics[width=1.0\columnwidth]{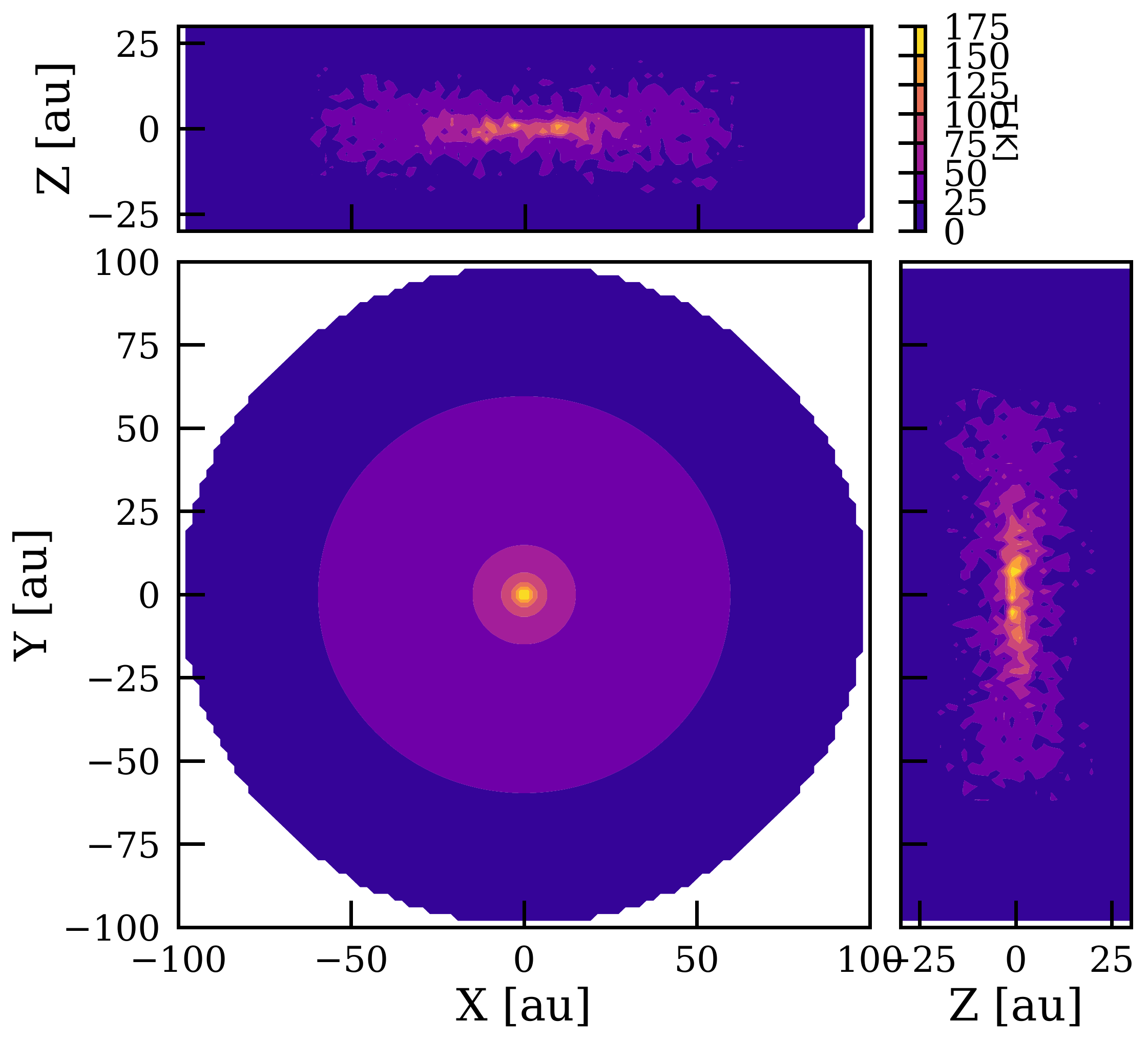}
\caption[]{Top view of the protoplanetary disk from three sides, the
  top view ($x$-$y$) image in the middle and two size aspects to the
  right and top. The colors represent temperature, and the coding is
  represented by the bar to the top right.
  \label{Fig:Disk_temperature}
}
\end{figure}

The mid-plane temperature profile for a disc in a steady state and
with constant viscosity follows a power law, $T_c \propto r^{-p}$, and
the surface density profile $\Sigma \propto r^{p-\frac{3}{2}}$ between
a few and $\sim 100$\,\AU\, \citep{1973A&A....24..337S}.  For
planet-forming disks in steady state $p \simeq 0.5$
\citep{2011ARA&A..49..195A} and surface density profile is roughly
$\propto r^{-1}$.  This profile is consistent with the surface
densities of proto planetary observed in the Ophiuchus star-forming
region \citep{2009ApJ...700.1502A,2010ApJ...723.1241A}.  We adopt a
power-law density profile with exponent $-1$ (with temperature profile
$\propto r^{-0.5}$) with Savronov-Toomre Q-parameter $q_{\rm out} =
25$ \citep{1960AnAp...23..979S,1964ApJ...139.1217T}, for which the
disk is everywhere at least marginally stable.  The temperature of
this disk ranges from 19\,K (at the rim) to about 165\,K, in the
central regions.  In Fig.\,\ref{Fig:Disk_temperature} we present the
disk temperature along three planes.  At $t=0$, when we start our
calculations, the supernova is ignited at a distance $d$ and with an
angle $\Theta$ with respect to the locus of the disk ($\Theta=0^\circ$
indicates right above the disk).  In Fig.\,\ref{Fig:geometry} we
present a schematic presentation of the simulation.  We performed
calculations for a grid in distance to the supernova $d = 0.05$\,pc,
0.1, 0.15, 0.2, 0.3, 0.4 and 0.6\,pc, and angle of the disk $\Theta =
15^\circ$ to 75$^\circ$ in steps of 15$^\circ$.

\subsection{The supernova lightcurve}

\begin{figure*}
\includegraphics[width=1.0\columnwidth]{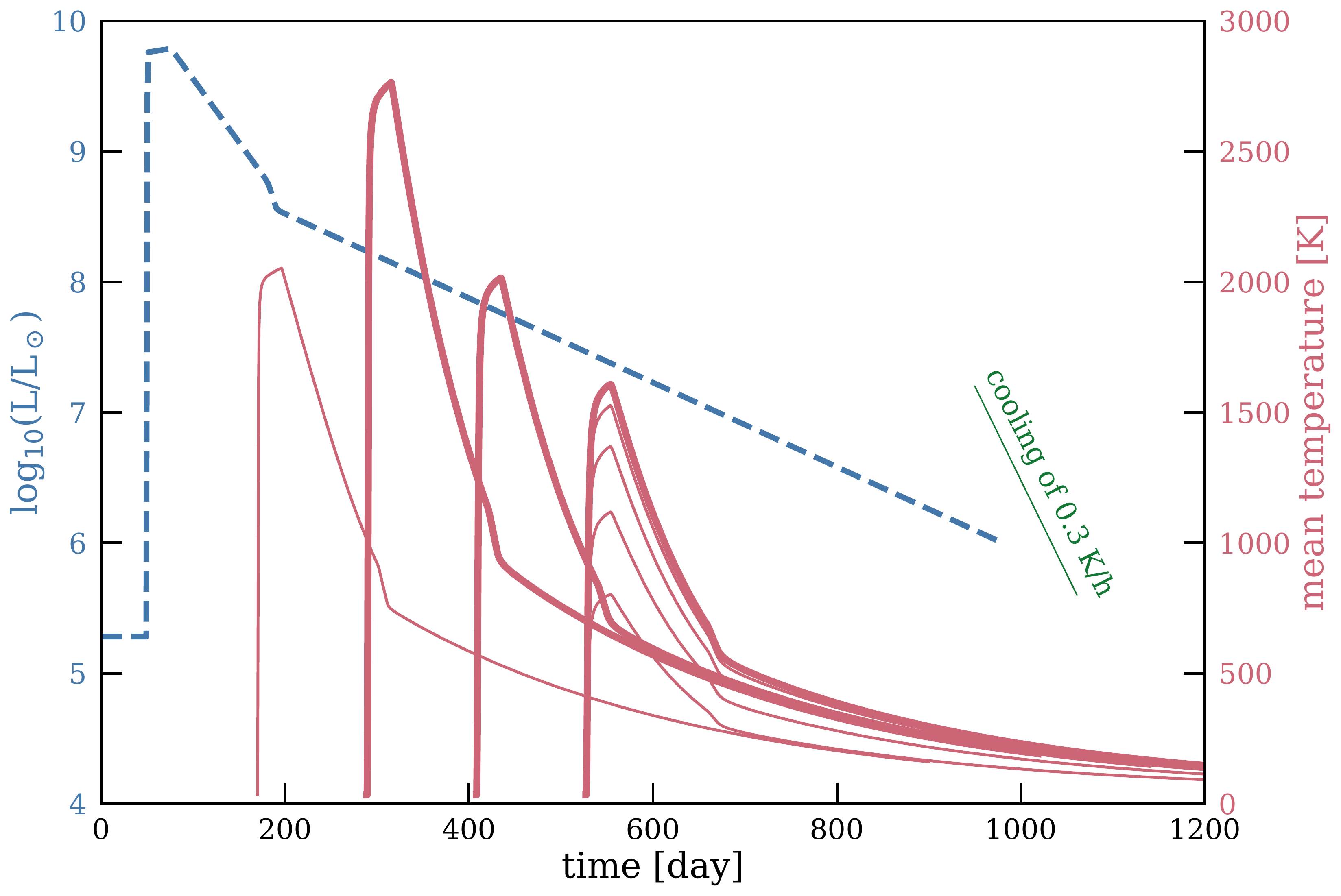}
~\includegraphics[width=1.0\columnwidth]{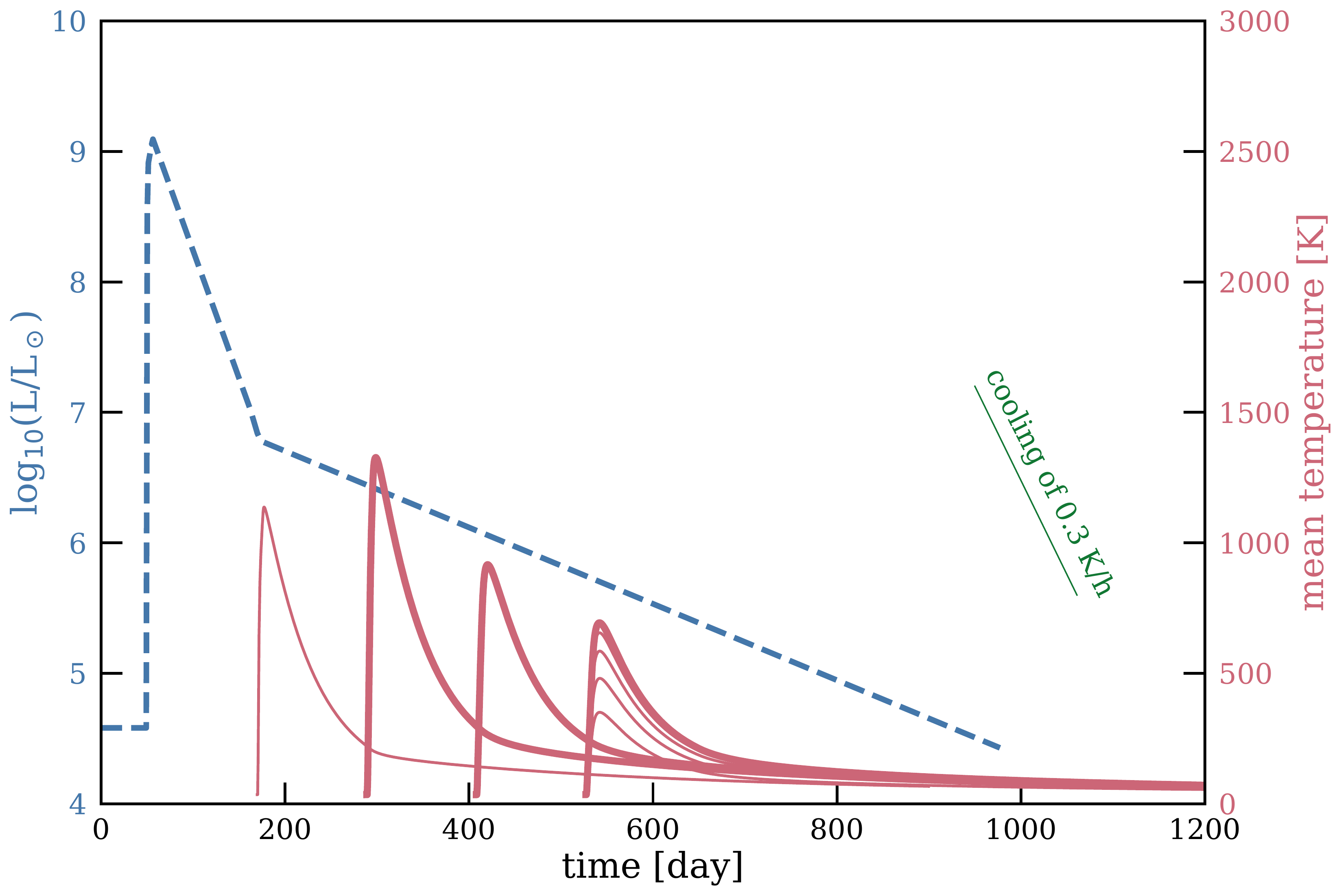}
\caption[]{Lightcurves and disk temperatures for the type IIp
  supernovae SN11aof (left panel) and SN10a (right panel).  The
  dashed blue curves with the scales in blue to the left of the
  respective panels give the luminosity of the supernova, are are
  power-law fits from \citep{2015ApJ...799..208S}. The red curves give
  the mean temperature of the disk at distances of 0.1, 0.2, 0.3 and
  0.4\,pc (from left to right). For the latter we present curves
  (bottom to top curve) for an inclination of $15^\circ$, $30^\circ$,
  $45^\circ$, $60^\circ$ and $75^\circ$. For the distance of 0.1\,pc
  we only show the disk temperature for the inclination angle of
  $75^\circ$, and for the other two distances at $15^\circ$.  In the
  right-hand panel, we show the same curves but for supernova SN10a.
  The thin green line (to the upper right) indicates a cooling of 0.3
  K/h, which is consistent with the cooling after the supernova
  radiation has heated the disk for SN11aof.  In
  \S\,\ref{Sect:heating} we discuss the effect of the heating on
  structure of vitreous chondrules in the disk.
  \label{Fig:SNLightcurves}
}
\end{figure*}

For the supernova light curve, we adopt the multiple power-law fits to
observed supernovae by \cite{2015ApJ...799..208S}.  We use fit
parameters for supernovae PS1-10a and PS1-11aof from their Table\,3,
because the lightcurves as well as the mass of the ejecta are well
constrained for these two.  For further reference, we call them SN10a
and SN11aof respectively.  The former peaks at a luminosity of $2.3
\times 10^{42}$\,erg/s (almost $10^{8.8}$\,\LSun) in about 10 days
after the supernova, whereas the latter reaches its peak of $1.1
\times 10^{43}$\,erg/s (almost $10^{9.8}$\,\LSun) about 26 days after
the supernova explosion is initiated.  We compare the light curves of
these two supernovae in Fig.\,\ref{Fig:SNLightcurves}.  From the fits
to the light curves, \cite{2015ApJ...799..208S} derive masses for the
ejecta of 6.0\,\MSun\, for SN10a and 23.5\,\MSun\, for
SN11aof.

The temperature of the supernova before hydrogen recombination in the
remnant is very high.  The temperature decays quickly with time and up
to about 100 days after the explosion follows the relation
\begin{equation}
  T(t) = T_0 \left( L(r)/L_{\rm peak} \right)^{\alpha}.
\end{equation}
Here $T_0 = 15,000$\,K and $\alpha =
1/3$\,\citep{2018MNRAS.473..513F}.  We adopt this temperature
evolution in our calculations.

\subsection{The combined hydrodynamics with radiative transfer solver}

The first part of the calculation is composed of a hydrodynamics
solver for the circumstellar disk and a radiative transfer code for
irradiating the disk.

The hydrodynamics is addressed with the smoothed particles
hydrodynamics code {\tt Fi}
\citep{1989ApJS...70..419H,1997A&A...325..972G,2004A&A...422...55P}.
We performed a series of test calculations for the hydrodynamics of
the protoplanetary disk using $10^{3}$ SPH particles, doubling the
number of particles to $10^{7}$ particles and these reached
convergence at a few times $10^4$ particles. Based on this validation
experiment we decided to perform all further calculations with $10^5$
SPH particles.

The radiative transfer calculations are performed with {\tt SPHRay}
\cite{2006PhRvE..74b6704R,2008MNRAS.386.1931A,2011ascl.soft03009A}. This
is a classic Monte-Carlo ray-tracing code meant for integrating rays
through the interpolated SPH density distribution.  The radiative
transfer code requires the ionizing luminosity of the radiative source
to be expressed in photons per second.  We adopt each photon to
deposits 20\,eV, which for SN11aof, which has a peak luminosity of
$10^{9.8}$\,\LSun\, results in $\sim 7.6\times 10^{56}$ photons/s. For
the energy spectrum of the photons, we assume a power-law index of
$-3$.  As soon as the outflow material blocks the central source the
spectrum becomes softer, but since we are mostly interested in the
hard and hot phase shortly after the onset of the supernova we ignore
this.  Absorption is computed from the transfer equation in each cell,
upon which re-emission creates new sources. The transfer equation is
solved using a semi-implicit iteration scheme
\citep{2006MNRAS.371.1057I}.  In \S\,\ref{Sect:Dust} we discuss the
effect of adding dust to the radiative transfer solution.

The two codes are combined using the Astronomical Multipurpose
Software Environment \citep[{\tt AMUSE} for
  short,][]{2011ascl.soft07007P,2013CoPhC.183..456P,2013A&A...557A..84P}.
AMUSE is a component library dedicated to multi-scale and
multi-physics astronomical calculations.  Details on the code can be
found in the associated textbook by \cite{AMUSE}.

The two codes are combined in the main event loop by taking subsequent
steps in the radiative transfer and hydrodynamics codes.  The loop
starts with half a step in the radiative transfer code.  This
calculation if followed by one full step in the hydrodynamics, and a
finishing second half-step in the radiative transfer.  During these
steps, the positions, velocities, densities, pressure, internal
energy, and degree of ionization are communicated between the
radiative transfer and the hydrodynamics solvers.  This coupling
strategy is called {\em interlaced temporal discretization} in the
{\tt AMUSE} framework \citep{AMUSE}, which is a form of operator
splitting \citep{1991AJ....102.1528W,2013A&A...557A..84P}.

We performed a series of test calculations with a step size of 1-day,
halving the step size by factors of two until we reached 10-minutes
time steps, for which convergence was reached at a step size of about
2 hours. Based on these validation experiments we decided to perform
all further calculations with a conservative step size of 1 hour.

We performed a series of radiative transfer test-calculations using
$10^5$ rays per time step (1 hour at a distance of 0.1\,pc), doubling
again until we reached $10^9$ rays per time step, and convergence was
reached at about $10^6$ rays per time step.  Based on this validation
experiment we decided to perform all further calculations with $10^7$
rays per time step.

The adopted backward difference solver for the combined evolution of
the temperature and abundances of chemical species includes photo
ionization, collisional ionization, recombination, photo heating and
atomic cooling \citep{2008MNRAS.386.1931A}.  The adopted chemical
model includes an approximation for the H2 chemistry taking into
account photo-ionization of the H$_2$ molecule. The frequency and flux
of the incoming radiation are calculated using the underlying
supernova model and assuming a thermalized spectrum.

The production calculations are performed on a 192-core X86-E7
workstation and took about a month per run.

\subsection{Results on the irradiation of the circumstellar disk}\label{Sect:RadiationTheDisk}

\begin{figure}
\includegraphics[width=1.0\columnwidth]{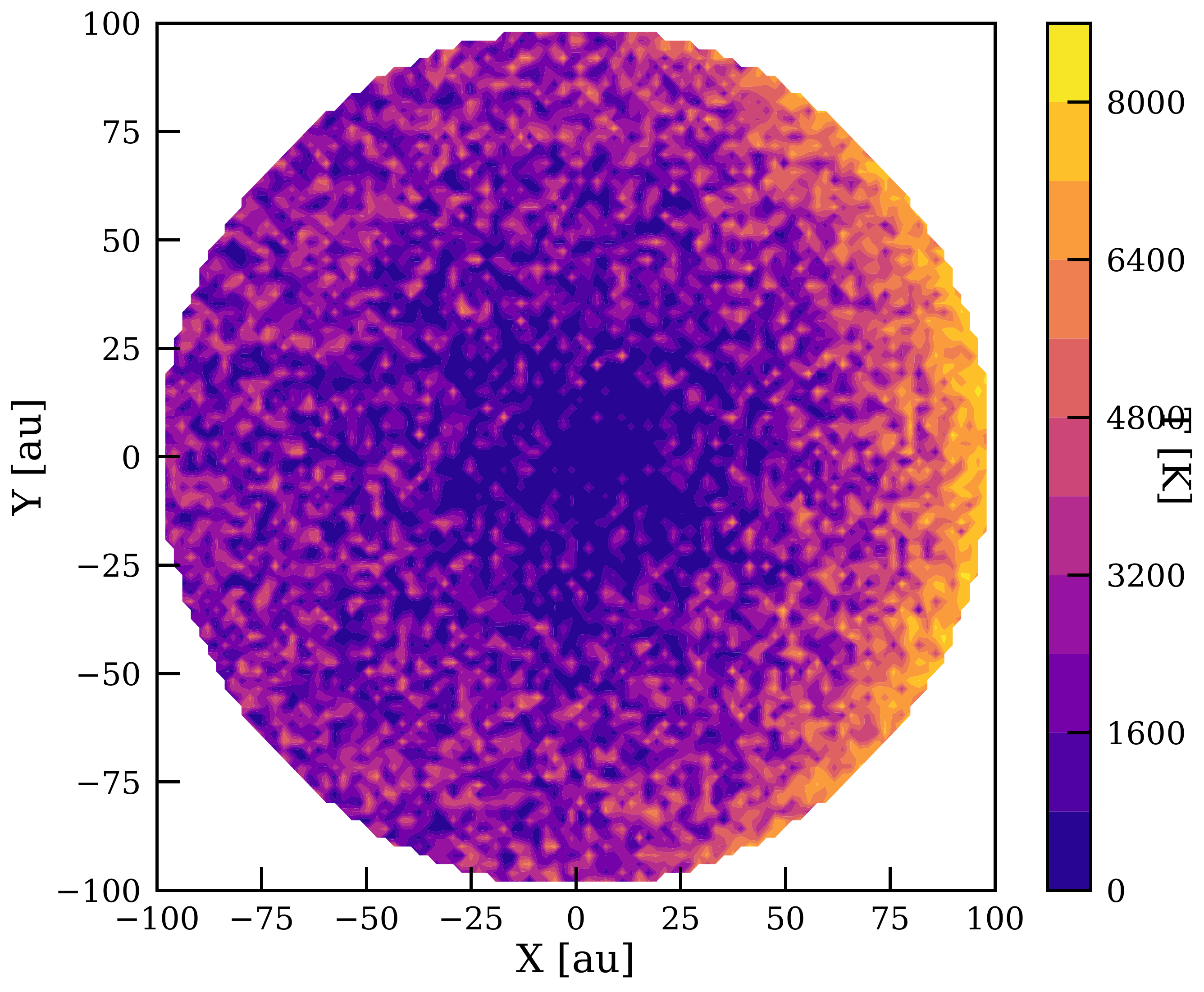}
\caption[]{View of the midplanet temperature of the circumstellar disk
  at the moment supernova SN11aof reaches its maximum. The supernova
  exploded at a distance of 0.2\,pc and at an angle of
  $\theta=60^\circ$. In fig.\,\ref{Fig:SN11aof_disk_temperature} we
  present several face on temperature distribution of the same disk at
  different moments in time.
  \label{Fig:Top_view_dusttemp}
}
\end{figure}

\begin{figure*}
  \begin{center}
  \includegraphics[width=0.4\columnwidth]{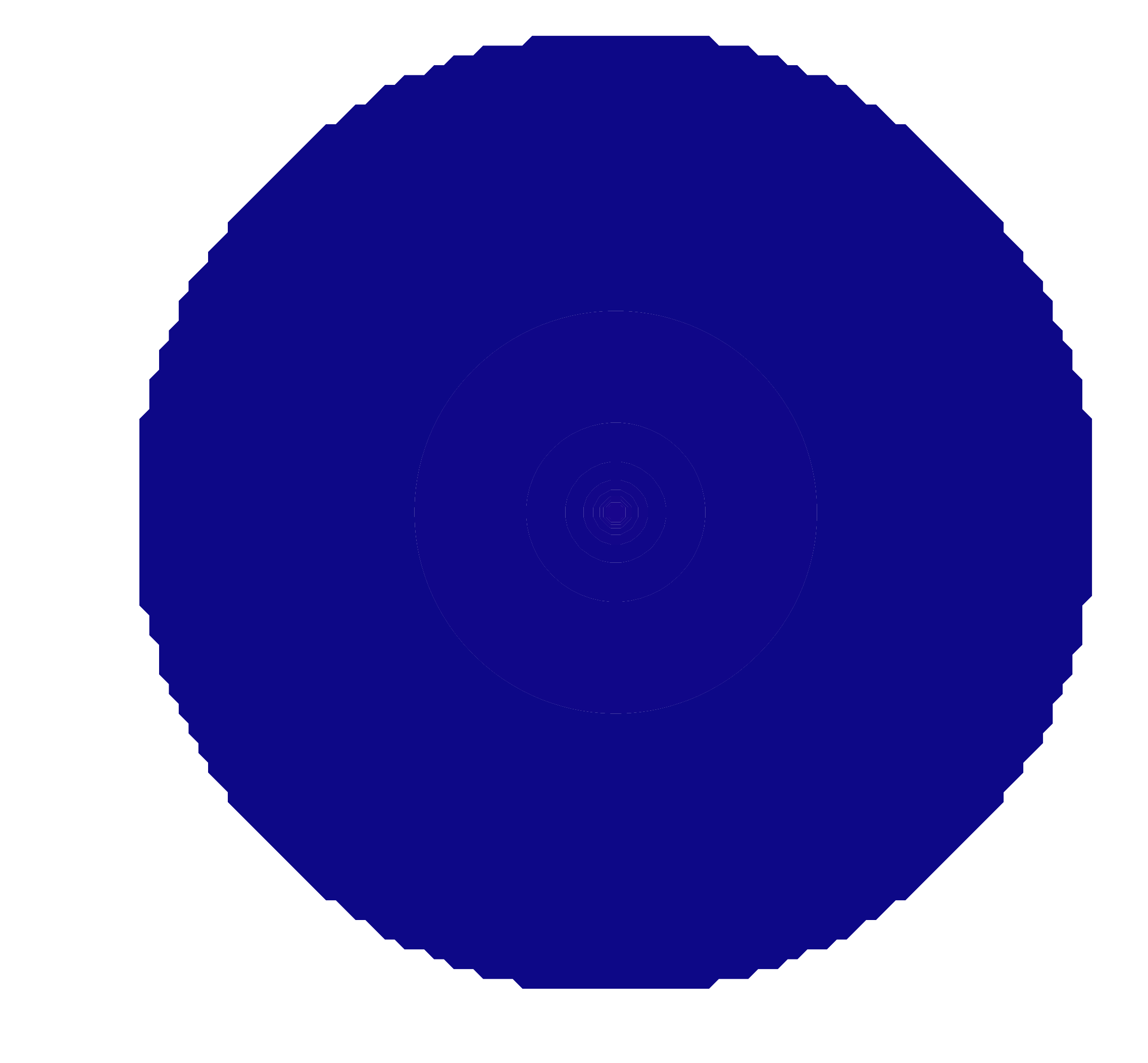}
~  \includegraphics[width=0.4\columnwidth]{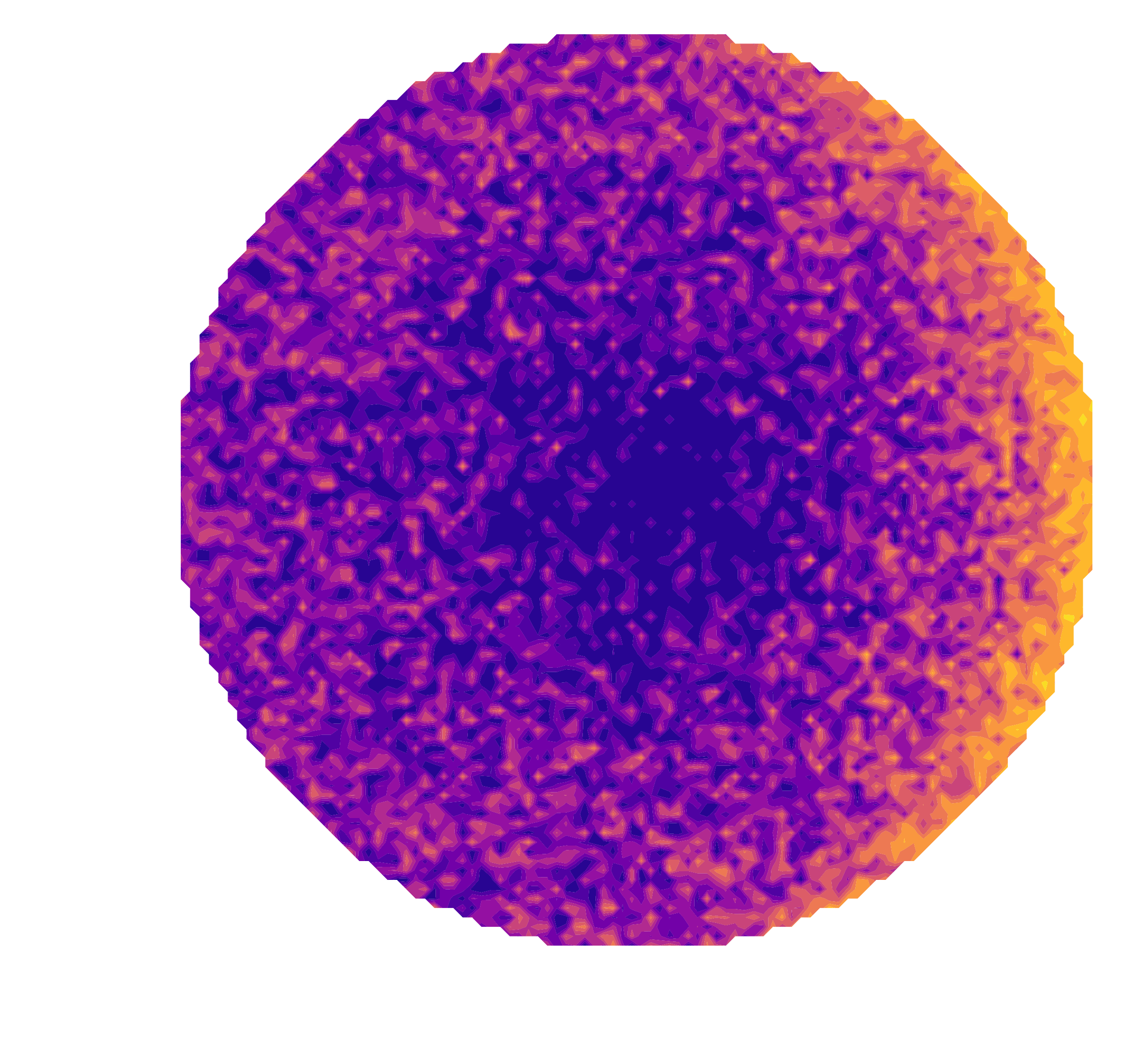}
~  \includegraphics[width=0.4\columnwidth]{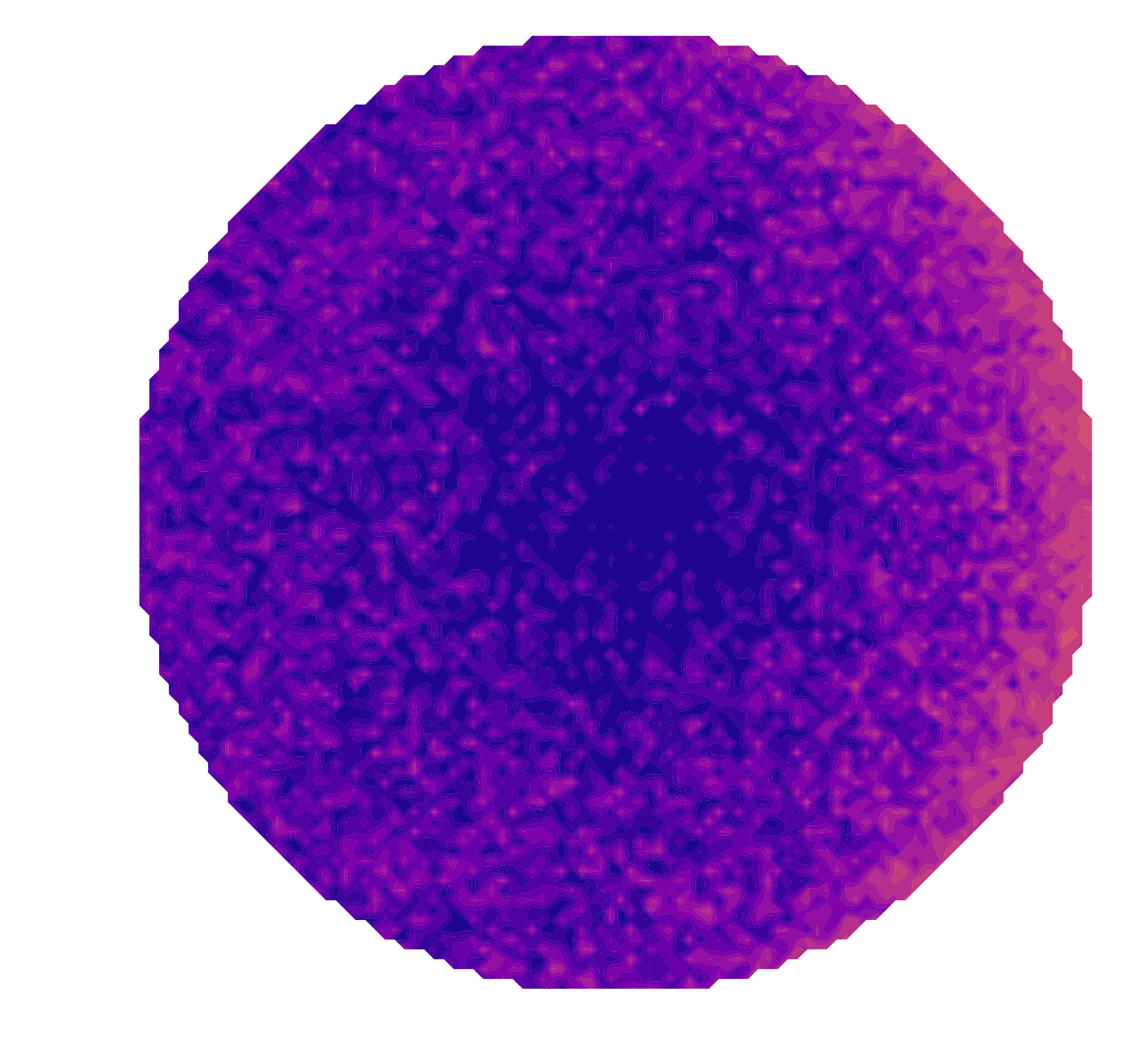}
~  \includegraphics[width=0.4\columnwidth]{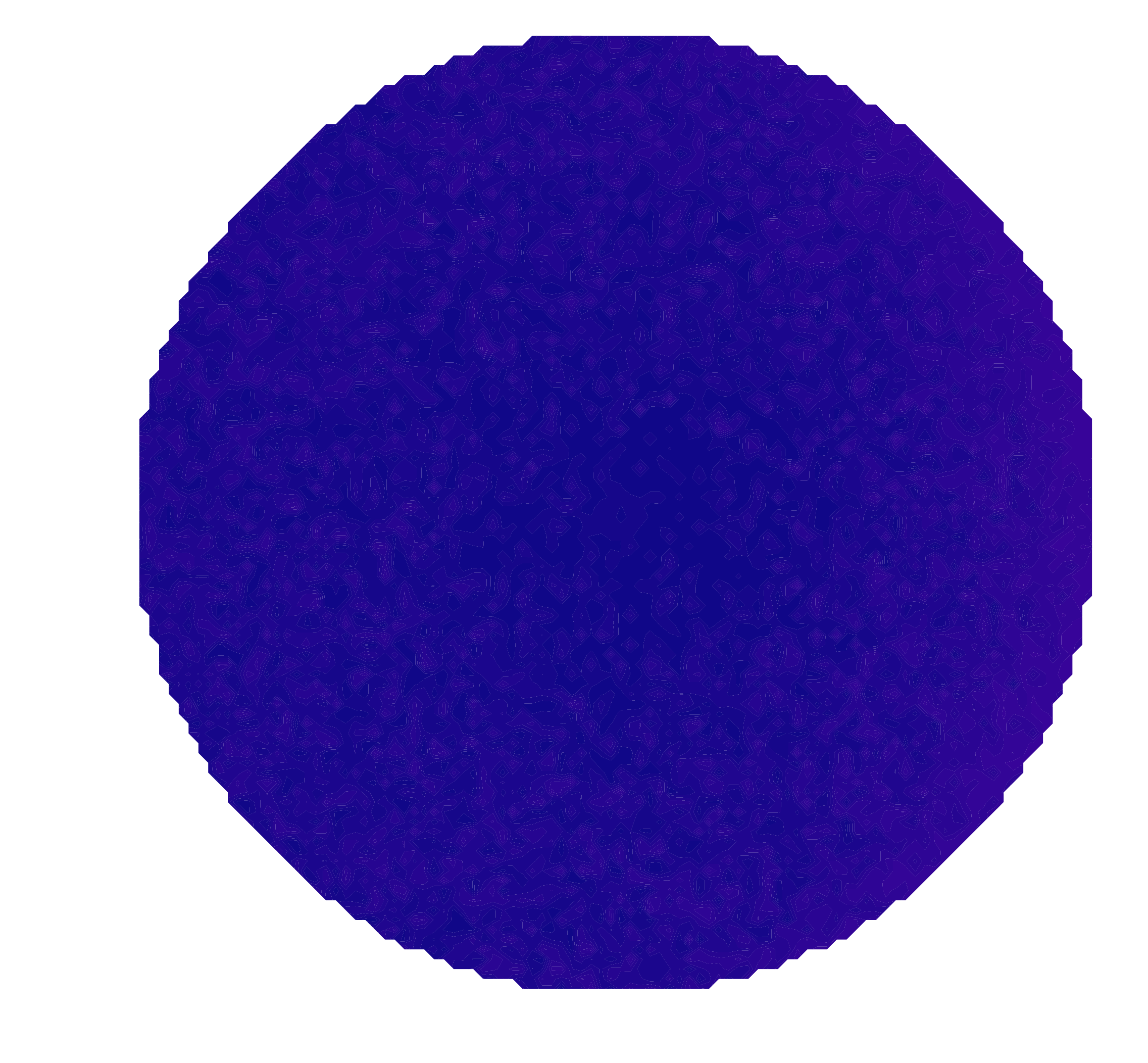}
\caption[]{Top view of the mid-plane temperature of the circumstellar
  disk at four epochs, the start of the simulation (left most image),
  at 26 days after the onset of supernova SN11aof, about 100 days
  after the supernova and about 2 years later, when the disk has
  almost completely cooled down again (right most image).  Each image
  was taken from the simulation where the supernova exploded at a
  distance of 0.2\,pc and at an angle of $\theta=60^\circ$.
  See Fig.\,\ref{Fig:Top_view_dusttemp} for the temperature scale.
  \label{Fig:SN11aof_disk_temperature}
}
\end{center}
\end{figure*}

The intense radiation of a bright supernova explosion increases the
average temperature in the disk.  In Fig.\,\ref{Fig:SNLightcurves} we
present the results of several calculations of the combined solver,
hydrodynamics plus radiative transfer.  We present various curves for
a range of angles and distances to both supernovae.  In
Fig.\,\ref{Fig:Top_view_dusttemp} we present a single top view of the
temperature in the disk's midplane as it is illuminated and heated by
supernova SN11aof from a distance of 0.2\,pc at an angle of
$\theta=60^\circ$.  In Fig.\,\ref{Fig:SN11aof_disk_temperature} we
present three more images of the same disk and supernova simulation at
different epochs.

\begin{figure*}
\begin{center}
\includegraphics[width=1.0\columnwidth]{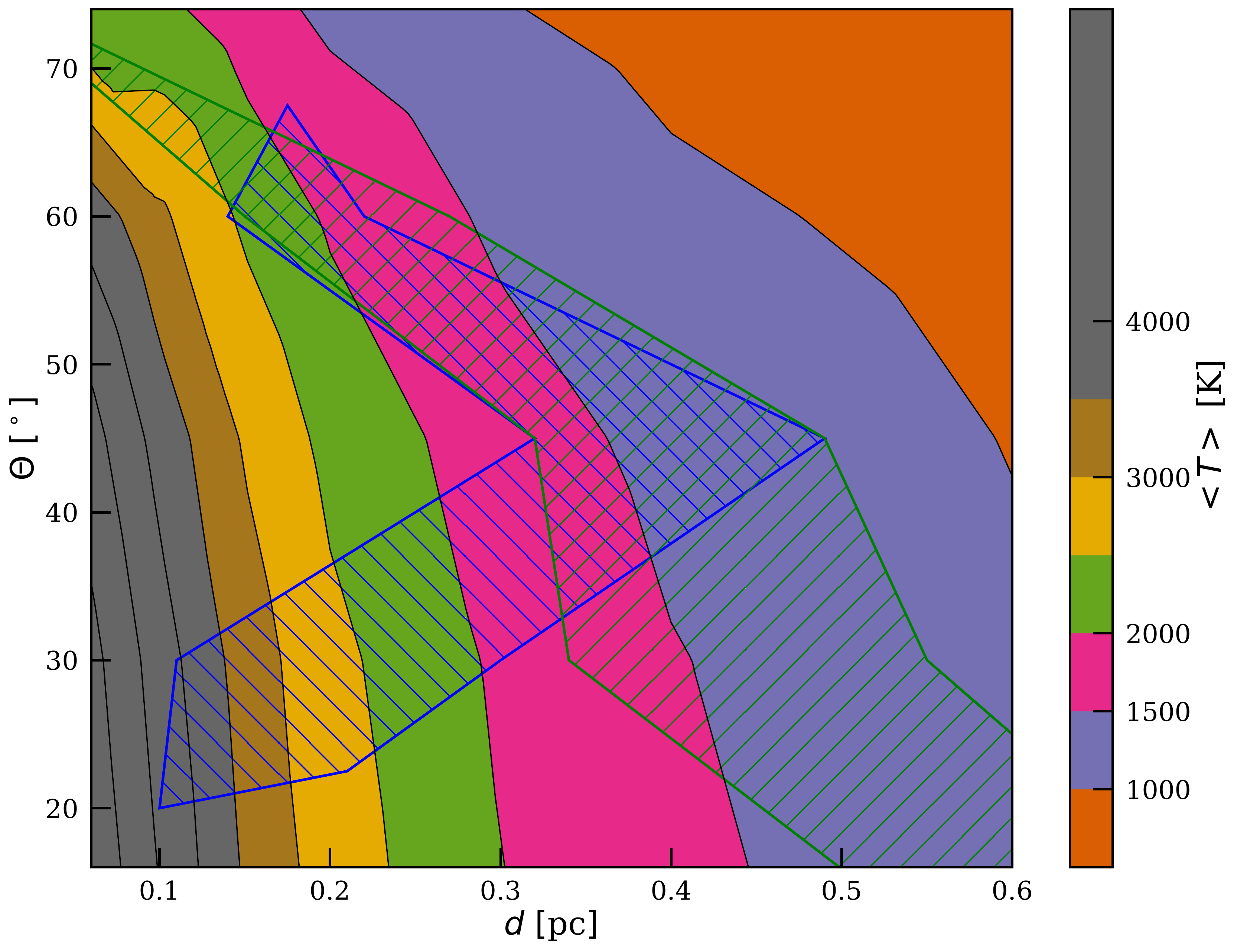}
~\includegraphics[width=1.0\columnwidth]{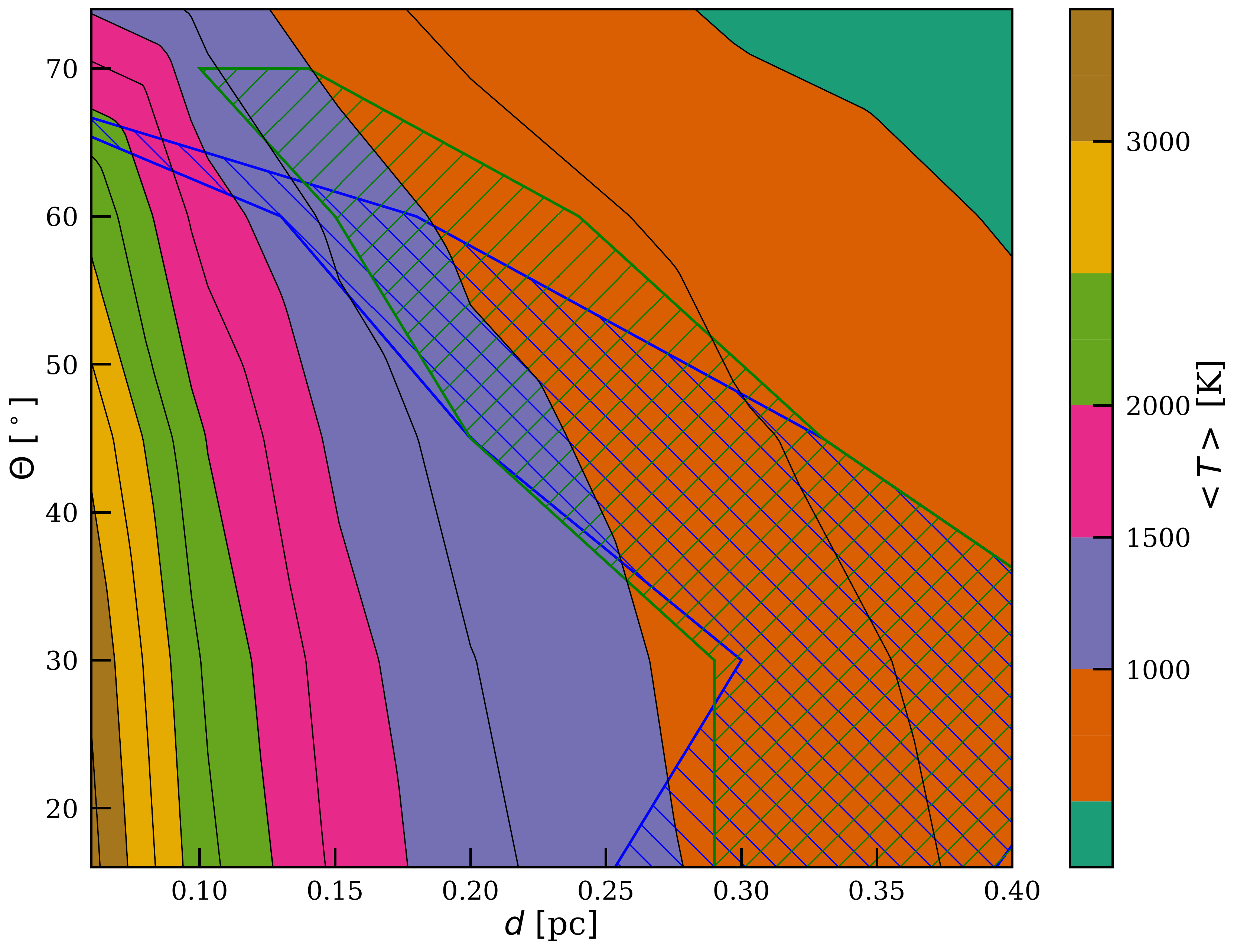}
\caption{Peak mean temperature of the disk (contours) due to the
  supernova radiation as a function of the distance to a supernova
  ($d$ in parsec) and the impact angle ($\Theta$ in degrees, measured
  from the disk's angular-momentum axis).  The adopted type IIp
  supernovae, SN11aof (left) has a peak luminosity of $1.1 \times
  10^{43}$\,erg/s and SN10a (right) has a peak luminosity of $2.3
  \times 10^{42}$\,erg/s.  Evaporation of the bulk of the solids in
  the proto-planetary disk is achieved when the irradiation heats it
  to a mean temperature of $\apgt 1500$\,K.  The two shaded regions
  indicate the range in distance for which the Sun's obliquity is
  consistent with the currently observed value of
  $5^\circ.6\pm1^\circ.2$ (blue with
  ``\textbackslash\textbackslash''\, hashes), and for which the disk
  is truncated to 42--55\,\AU\, (green with ``//'' hashes).  This area
  is indicated in Fig.\,\ref{Fig:disk_heatedfraction} and
  Fig.\,\ref{Fig:disk_enrichment} with a blue polygon.  The most
  promising part of parameter space is where the two hashed areas
  overlap because for there both criteria are met.
  \label{Fig:temperatureSNPS11aof}
  \label{Fig:temperatureSNPS10a}
}
\end{center}
\end{figure*}

Temperatures in excess of $\sim 1200$\,K are is sufficient to melt
chondrules without vaporizing them \citep{2005mcp..book..407D}.  In
Fig.\,\ref{Fig:temperatureSNPS11aof} (see also
Fig.\,\ref{Fig:temperatureSNPS11aof_with_dust} where we included the
effect of dust) we present the temperature of the disk when it is
hottest as a function of the distance to the supernova $d$ and the
impact angle $\Theta$.  In our simulations, temperatures of
1500--2000\,K are reached when the supernova exploded at a distance of
0.13--0.44\,pc for SN11aof and $\sim 0.05$--0.18\,pc for SN10a (for a
discussion on the effect of dust in the circumstellar disk see
\S\,\ref{Sect:Dust}).  The dependence of the peak mean temperature on
the inclination angle $\Theta$ is not very strong unless the
irradiation happens almost edge on ($\Theta \apgt 70^\circ$) when
self-shielding effects become strong.  The disk would vaporize if the
temperature would be even higher, for which there is no observational
evidence.

After the main irradiation, when the supernova starts to dim, the
circum-stellar disk material cools down to the equilibrium state. The
cooling rate in this phase is of the order of 0.3\,K/h (see the green
line to the right in Fig.\,\ref{Fig:SNLightcurves}).  After $\sim 3$
years, the radiation of the supernova has dropped below $\sim
10^5$\,\LSun\, at which moment we stop the radiative transfer
calculation.  The further thermal evolution of the disk is addressed
by integrating the thermal evolution using a semi-analytic
approximation for the cooling curve \citep{1997A&A...325..972G}.

\section{The arrival of the blast wave}\label{Sect:blastwave}

\begin{figure}
\includegraphics[width=1.0\columnwidth]{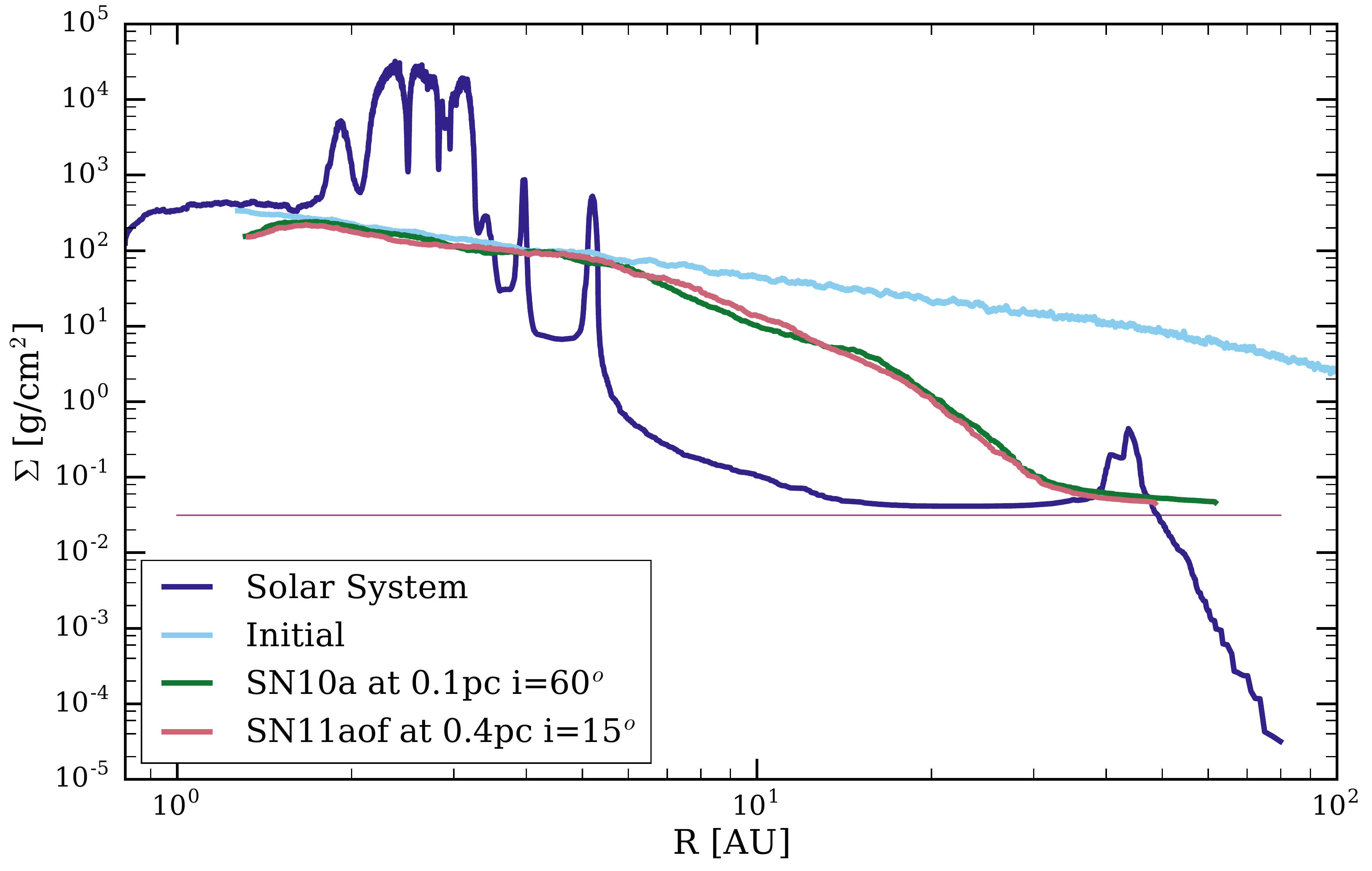}
\caption[]{Surface density profile of the proto planetary disk at two
  moments in time for two supernovae SN10a (green) and the much
  brighter SN11aof (red). The former was at a distance of
  $d=0.1$\,pc and with an impact angle of $\Theta = 60^\circ$, the
  latter at a distance of $d=0.4$\,pc and with an angle of $\Theta =
  15^\circ$.  The dark blue curve indicates the surface density
  distribution of the known planetesimals in the Solar System (see
  {\tt http://minorplanetcenter.net/data}).  The total mass of the
  Sun's planetesimal disk was normalized to the initial disk mass in
  our simulations (0.01\,\MSun). The initial density profile of the
  disk (light blue) is the same for each model. The thin
  periwinkle-colored horizontal line indicates the minimum disk-mass we
  adopted, truncating the Solar System's disk at about 55\,\AU.
  \label{Fig:SNcomparison_diskprofile}
}
\end{figure}

The outflow velocity of the two supernovae is $\sim 10^4$\,km/s for
SN10a and 5400\,km/s for SN11aof.  At the
short distance ($\aplt 1$\,pc) this outflow is still in the free
expansion phase, and the blast wave hits, some $10$--$100$\,yr after
the initiation of the supernova.  By that time the disk has cooled
down to $T_{\rm disk} \sim 100$\,K. The morphology of the disk is
hardly affected by the irradiation of the supernova (see
Fig.\,\ref{Fig:SNcomparison_diskprofile}) because the heating and
cooling by the supernova radiation proceeded so quickly.

Instead of following the entire supernova blast wave, which would be
difficult for performance arguments and numerical stability, we
initiate the hot gas at a distance of 10\,\AU\, from the nearest point in
the circum-stellar disk. With this adopted distance and the velocity
of the blast wave, it takes 3--100 days before the shock front of the
supernova traverses the entire disk, depending on the impact angle and
the supernova blast-wave velocity.

The hot shell is initialized as a cylinder directed at the Solar
System. This cylinder has a radius of $r_{\rm shell} = 150$\,\AU, to
assure that the entire circumstellar disk is engulfed. The mass along
the shell is distributed homogeneously.  Along the cylinder axis, we
adopt a density profile with the peak nearest to the Solar System
dropping-off linearly in $log$ density with a characteristic scale of
200\,\AU\, along the far side. The total mass in the cylinder is
\begin{equation}
  m_{\rm shell} = m_{\rm eject} \left( {r_{\rm shell} \over 2 d} \right)^2.
\end{equation}
As a consequence, the mass and density of the impacting shell of SN10a
at a distance of $d=0.1$\,pc, is comparable to that of SN11aof at
a distance of $d=0.2$\,pc.

The temperature of the shell at a distance of 10\,\AU\, from the
exploding star is $T_{\rm eject} = 9 \times 10^7$\,K, but by the time
the gas reaches the Solar System it has cooled to
\citep{1990ApJ...356..549S}
\begin{equation}
  T_{\rm shell} \propto T_{\rm eject} \left( {d \over 10 {\rm \AU\,}} \right)^{-3(\gamma-1)},
\end{equation}
which for $d = 0.1$\,pc results in $T_{\rm shell} \simeq 9500$\,K, and
to about 1800\,K at a distance of $d=0.4$\,pc.  Here we adopted
$\gamma = 7/5$.  The sound speed in the shell then is about $c_s
\simeq 5.8$\,km/s, which is negligible compared to the speed of the
shell. As a consequence, the shape of the shell front remains intact
if it would traverse the disk without obstruction.

\subsection{The effect of the supernova blast wave on the disk}

The supernova outflow of hot gas hits the disk with an angle $\Theta$
with respect to the rotation axis of the disk (see
Fig.\,\ref{Fig:geometry}).  The blast wave takes a few weeks to pass,
during which the disk heats again to high temperatures, it harasses
the disk and injects small amounts of blast wave material. Each of
these processes has profound consequences for the disk structure and
further evolution.
About 100\,days after first contact between the
disk an the supernova shell the disk mass reaches its final
value. After this moment the temperature and density profile of the
disk hardly change any further.  We check this by continuing several
calculations to 4\,yr after the shell impact, but most calculations
are stopped around 200 days after first contact. Further analysis is
performed by averaging over the last 10 snapshots of the simulations,
each 1\,day apart.

The supernova blast-wave injects energy into the disk. This excess
energy results in its heating and ablation, the effect of which we
will discuss in the following \S. The amount of energy injected
depends on the distance to the supernova and the incident angle. The
amount of kinetic energy injected into the disk is generally about an
order of magnitude smaller than the amount of kinetic energy lost by
the blast wave material. The amount of energy gained by the disk varies
between and ranges from $3.0 \times 10^{45}$\,erg at 0.1\,pc to $2.8
\times 10^{43}$\,erg at 0.6\,pc, and follows $\Delta E_{\rm kin}
\simeq 10^{42.5 - 3u}$\,erg, with $u = \log_{10}(d/{\rm pc})$.

\subsubsection{Disk stripping}

Ram-pressure stripping by the blast wave has a strong effect on the
size of the disk \citep[see also][]{2016A&A...594A..30W}.  If the
supernova occurs within $\sim 0.05$\,pc most of the disk is
obliterated.  We consider these results not representative for the
Solar System because this is hardly reconcilable with the current size
of the planetary system and the edge of the Kuiper belt.  But at a
distance of 0.15--0.4\,pc the disk is truncated at about $42$ to
$55$\,\AU\, almost irrespective of the supernova, although there is a
strong correlation with the impact angle $\Theta$. Such a disk size is
consistent with the current extend of the Kuiper belt.  In
Fig.\,\ref{Fig:temperatureSNPS11aof} we indicate the area for which
the circumstellar disk was truncated in these limits by the green ``//''
hashed area.

In Fig.\,\ref{Fig:SNcomparison_diskprofile} we present the surface
density profile of the disk that is hit by the blast wave at two
moments in time; the moment the blast wave hits the disk (light blue
curve) and 4 years after the impact. The latter is presented for two
models, one for SN10a at a distance of 0.1\,pc and with an impact
angle 60$^\circ$ (rather edge-on), and for SN11aof at a distance
of 0.4\,pc and with an impact angle 15$^\circ$ (almost face-on).

After such an extended harassment by the supernova blast wave, the
outer parts of the disk have been stripped, up to a distance of about
50\,\AU\, from the host star. The disk between $\sim 10$\,\AU\, and
$\sim 50$\,\AU\, is depleted whereas within the inner $\sim 10$\,\AU\,
the disk is hardly affected.

For comparison, we present an estimate of the density profile of the
Solar System's planetesimals disk (dark blue). Here we simply assume
that each planetesimal contributes equally to the local density and
that the Solar System's disk has a total mass of 0.01\,\MSun,
equivalent to the initially adopted mass of the circumstellar disk.

\subsubsection{Tilting the disk}

The disk is not deformed uniformly by the passing blast wave, but the
periphery and the part of the disk closest to the supernova are more
strongly affected than the inner parts.  This interaction between the
supernova blast-wave and the disk material causes the latter to tilt,
orienting itself perpendicular to the flow due to the interaction
between the momentum of the gas and the inclined disk
\citep{2017A&A...604A..88W}.

For each simulation, we calculate the inclination of the disk with
respect to the initial inclination for each particle. We adopt the
median of the inclination of all particles bound to the disk as the
global disk inclination and a radial profile is determined by
calculating the median inclination in a co-moving bin of 500 particles.

\begin{figure}
\begin{center}
\includegraphics[width=1.0\columnwidth]{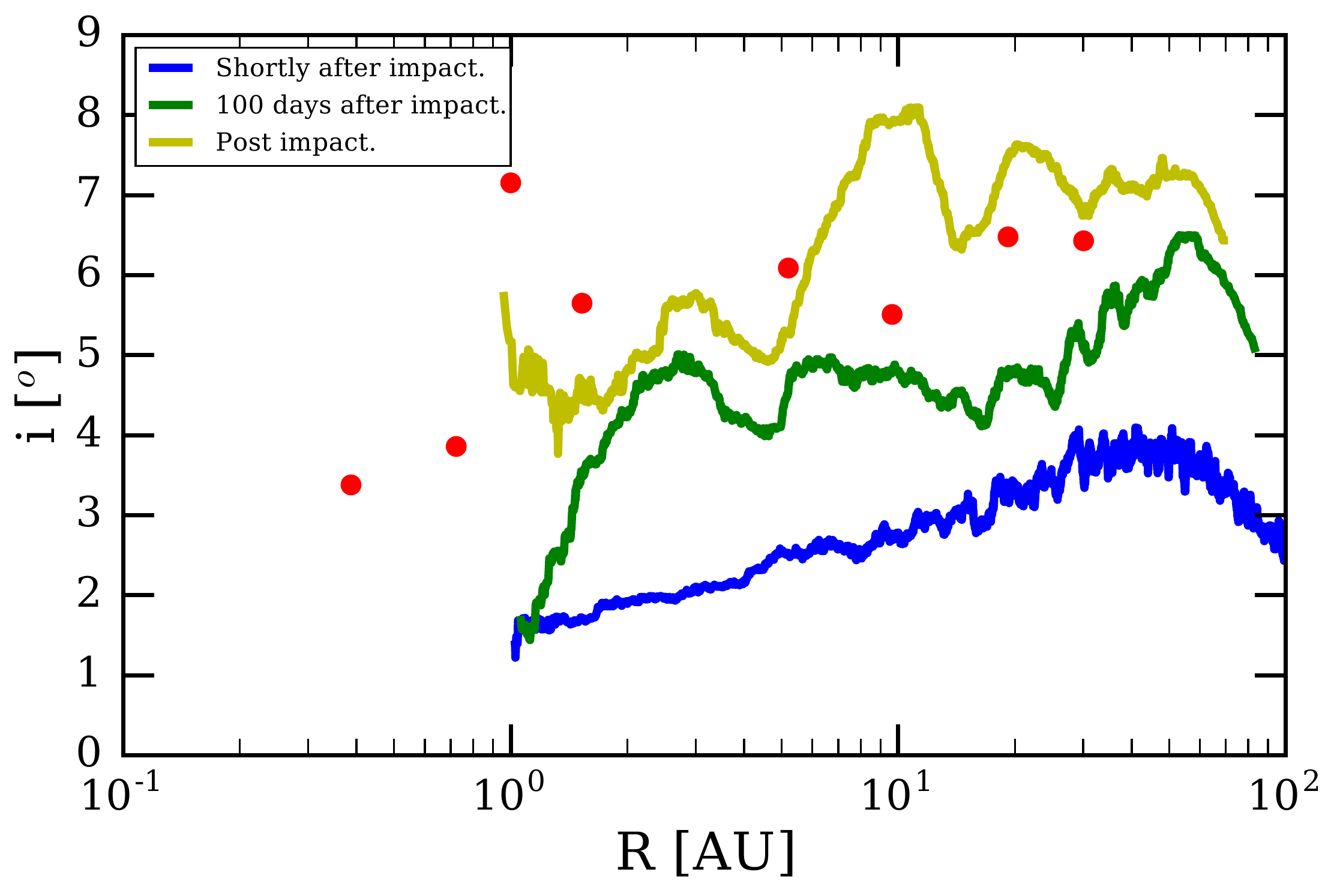}
\end{center}
\caption{Radial distribution of the inclination of particles bound to
  the disk shortly after moment of the supernova blast-wave impacts
  (blue), 100 days after the supernova shell hits the disk (green) and
  at the end of the simulation (yellow) for simulation of SN11aof
  at a distance of $d=0.4$\,pc and $\theta = 45^\circ$.  The bullet
  points give the inclinations of the 8 known planets.
  \label{Fig:radial_disk_inclination}
}
\end{figure}

The initial radial distribution in inclinations is rather flat and
near $i \simeq 0^\circ$.  At a later time, the distribution of orbital
inclinations increases, with higher inclination at larger distance
from the central star.  We demonstrate this in
Fig.\,\ref{Fig:radial_disk_inclination} where we present the radial
distribution of the inclination for the simulation with supernova
SN11aof at a distance of $d=0.4$\,pc and at an angle of $\Theta =
45^\circ$ at three moments in time.  The trend of an increased
inclination with disk radius is general for the simulations, and in
this example matches reasonably well with the inclinations observed in
the Solar System (see also Fig.\,\ref{Fig:radial_disk_inclination}).
The radial increase is expected from a theoretical perspective because
the time scale for tilting the disk is shorter at larger distance from
the star \citep{2017A&A...604A..88W}.  \cite{2017A&A...604A..88W} also
suggested that a nearby supernova could have this effect on the Sun's
protoplanetary disk, and could potentially explain the observed angle
between the Ecliptic and the Sun's equatorial plane of $\sim 5^\circ.6
\pm 1.2$ \citep{2005ApJ...621L.153B}.  In
Fig.\,\ref{Fig:temperatureSNPS11aof} we indicate this range with the
blue hashed area.

\begin{figure}
\begin{center}
\includegraphics[width=1.0\columnwidth]{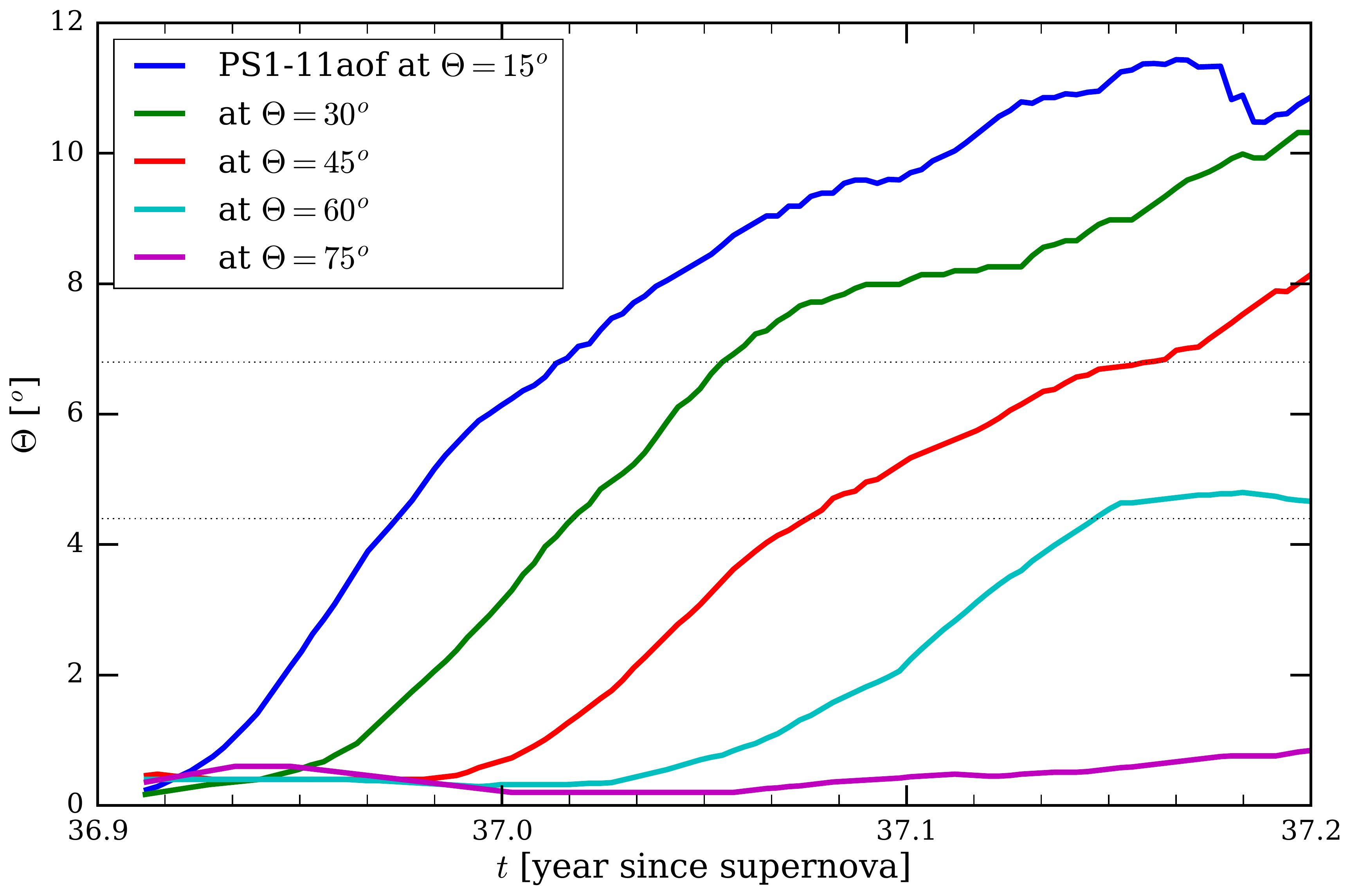}
\end{center}
\caption{The median inclination of the protoplanetary disk as a
  function of time since the supernova explosion of supernova
  SN11aof at a distance of 0.2\,pc for a range of impact angles
  of 15$^\circ$ to 75$^\circ$. The two horizontal dotted lines
  indicate the dispersion along the mean inclination of the current
  Solar System with respect to the Sun's equatorial plane.
  \label{Fig:disk_inclination}
}
\end{figure}

In Fig.\,\ref{Fig:disk_inclination} we present the evolution of the
inclination of the disk as a function of time from the moment the
supernova blast-wave hits the outer disk edge (about 36.9\,yr after
the supernova) up to the moment that the blast wave has completely
passed the disk (about 100 days later).  The presented calculations
are for a distance of $d=0.2$\,pc from supernova SN11aof.  When
this supernova occurred at an impact angle $\Theta \aplt 45^\circ$,
the final tilt of the disk exceeds the observed obliquity of the Sun.
The solution here must be degenerate. It is evident from
Fig.\,\ref{Fig:disk_inclination} that shallow impact angles are more
effective in tilting the disk.  The time scale on which eccentricity,
and therefore also the inclination, changes $\propto 1/(\sin(\Theta)
\cos(\Theta))$, which has a minimum at $45^\circ$.  For larger
inclinations, the flow is less effective in increasing the momentum
due to the smaller projected surface area of the disk with respect to
the incoming flow, while for smaller inclinations the force in the
plane of the disk that increases the eccentricity is smaller.
However, in order to precisely match the observed Sun's obliquity of
$\sim 5.6^\circ$ would require the supernova impact angle to be of the
same order, which leaves a very small range of possible impact
angles. We did, however, not consider the effect of very small $\Theta
\aplt 10^\circ$ supernova impact angles.

When the supernova occurs with a high impact angle of $\Theta \apgt
65^\circ$, the disk has no time to reorient itself into the supernova
blast-wave. In these cases and at a distance of 0.2\,pc from the
supernova the disk does not tilt to the observed value.  Only for an
angle of $\sim 60^\circ$ to $\sim 50^\circ$ the disk tilt is
consistent with the observed value of $5.6\pm1.2^\circ$ (indicated by
the horizontal dotted lines in Fig.\,\ref{Fig:disk_inclination}).

\subsubsection{Heating the disk}

\begin{figure*}
\begin{center}
\includegraphics[width=1.0\columnwidth]{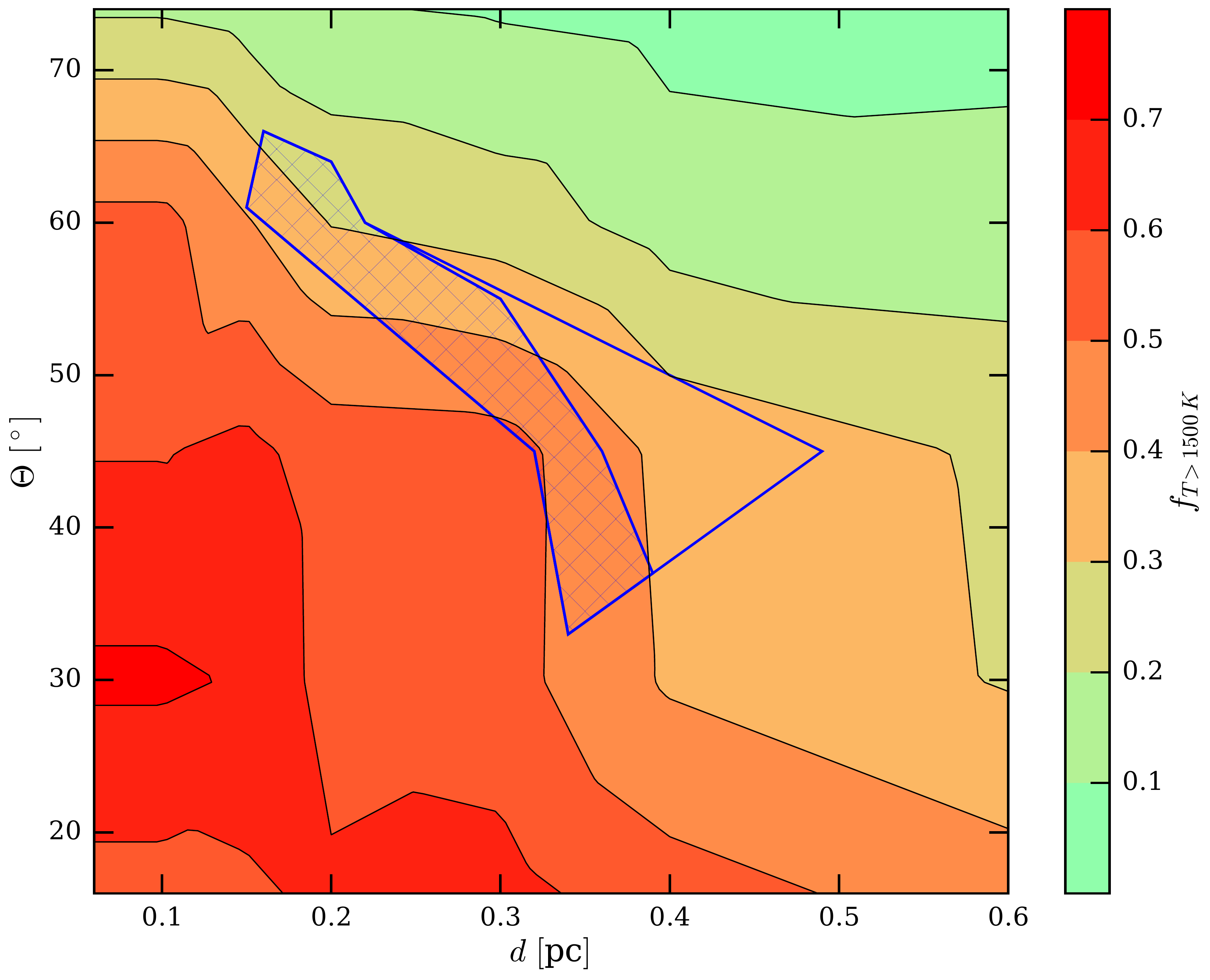}
~\includegraphics[width=1.0\columnwidth]{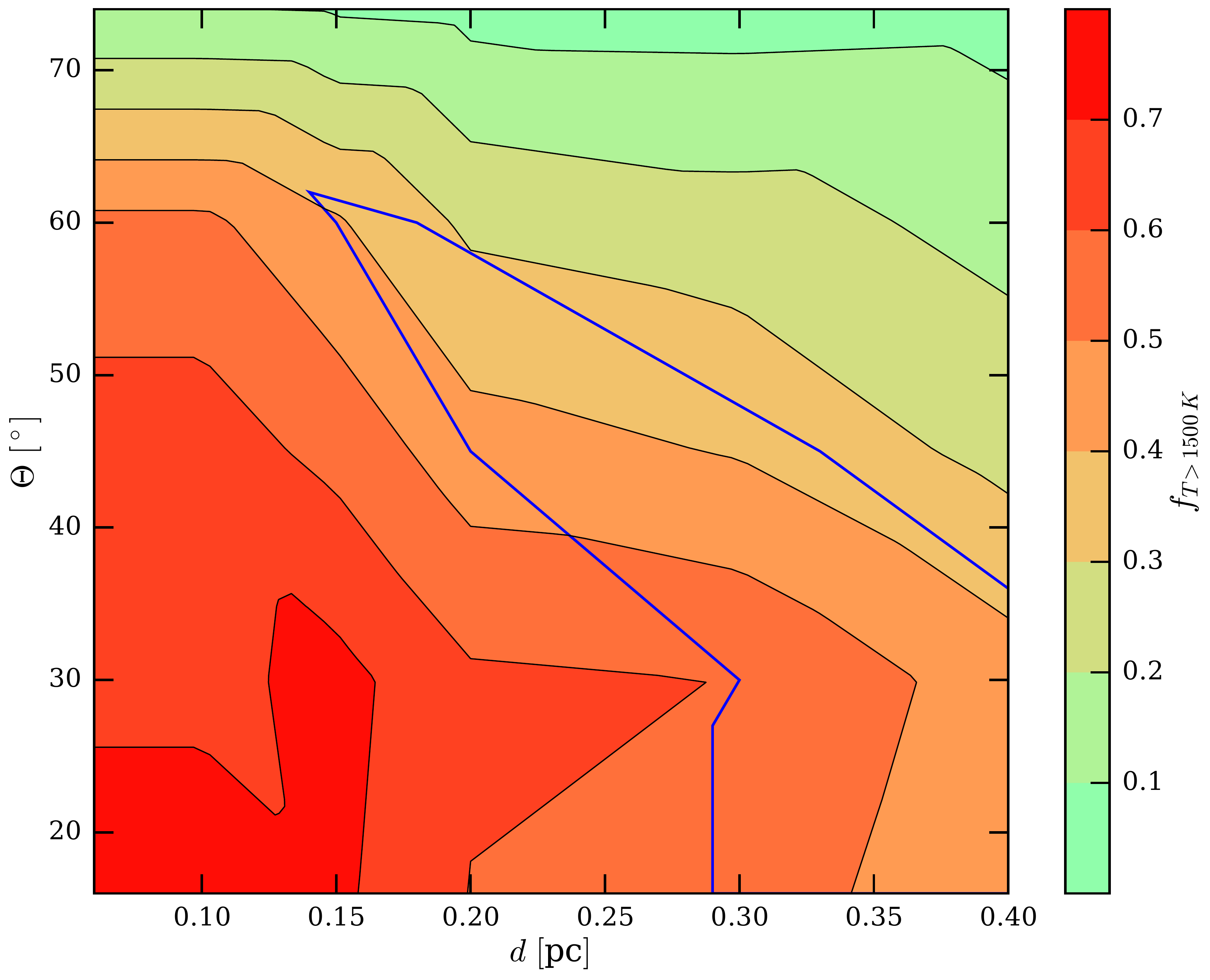}
\caption{Fraction of the surviving disk that is heated to a
  temperature $\apgt 1500$\,K by the supernova blast wave (scale to the
  right) for SN11aof (left) and SN10a (right).
  The enclosed area indicates the initial parameter space for
  which the observed obliquity and the disk size are satisfactory
  reproduced by the simulations (see
  Fig.\,\ref{Fig:temperatureSNPS11aof}). The hashed subset
  of this area indicates for which parameters the disk is also heated
  to a temperature $\apgt 1500$\,K by the supernova radiation. The
  non-hashed area to the lower right indicates for which parameters
  the disk is sufficiently truncated and tilted by the blast wave but
  the supernova flash-induced temperature is insufficient to melt
  corundum.
  \label{Fig:disk_heatedfraction}
}
\end{center}
\end{figure*}

With a temperature of $T_{\rm shell} \simeq 9500$\,K the blast wave
material is extremely hot when it interacts with the disk. As a
result, the disk itself will also be heated to high temperatures,
although this heating is not very efficient due to the low density of
the blast wave.  In Fig.\,\ref{Fig:disk_heatedfraction} we present the
mass-fraction of the disk that is heated to $\apgt 1500$\,K by the
supernova blast-wave.  We also present the overlapping regions where
the disk is truncated to $42$--$55$\,\AU\, and tilted to
$5.6\pm1.2^\circ$. Both supernovae have such an overlapping area in the
initial parameter space.  The blue-hashed area in the left panel of
Fig.\,\ref{Fig:disk_heatedfraction} indicates where the temperature
induced by the supernova radiation exceeds 1500\,K. Only supernova
SN11aof is able to heat the disk to this range.  In the overlapping
regions, a fraction of $0.3$--$0.4$ of the disk is heated to
sufficiently high temperatures to melt.

\subsubsection{Accretion from the blast wave}

\begin{figure*}
\begin{center}
\includegraphics[width=1.0\columnwidth]{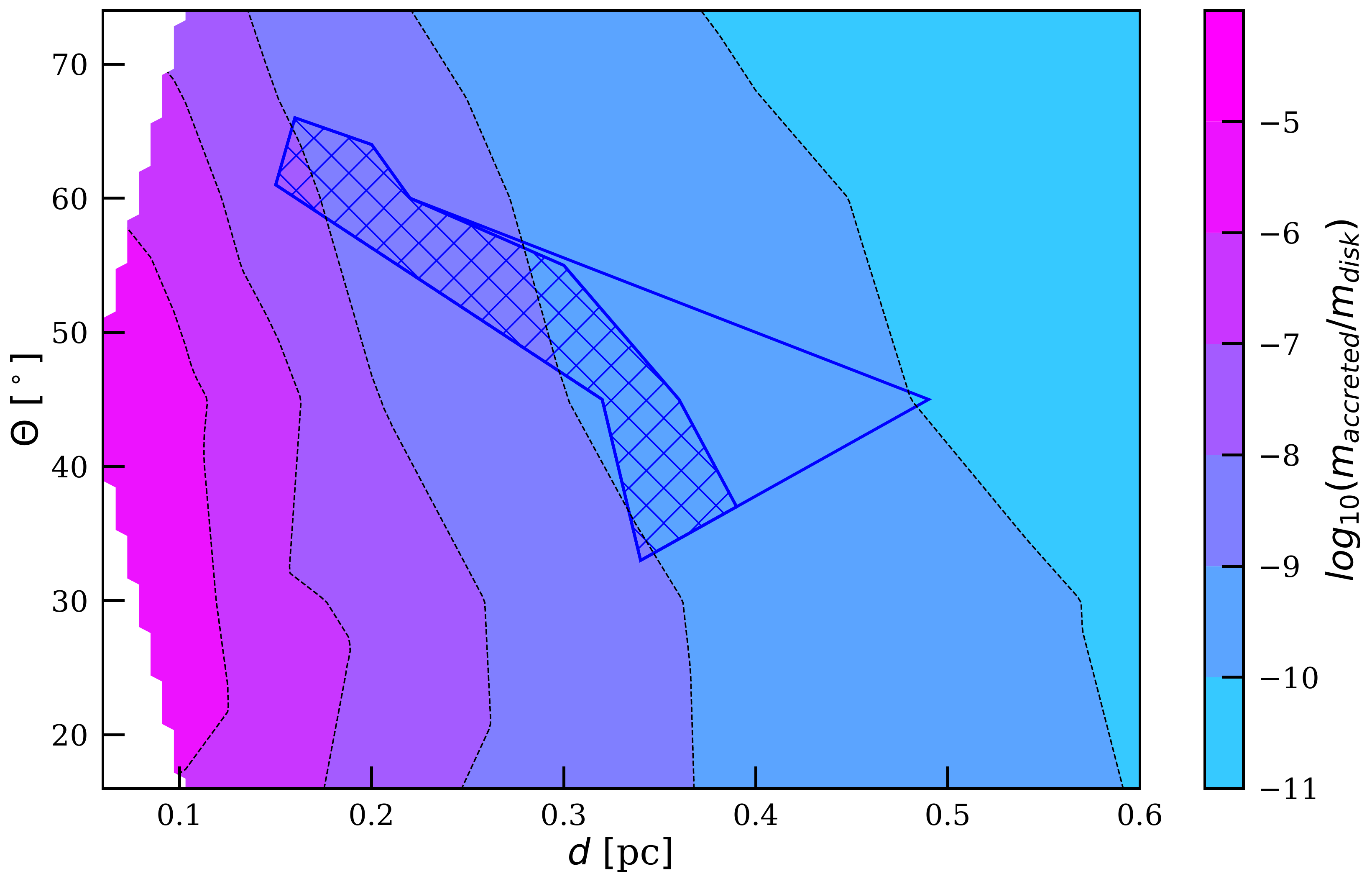}
~\includegraphics[width=0.93\columnwidth]{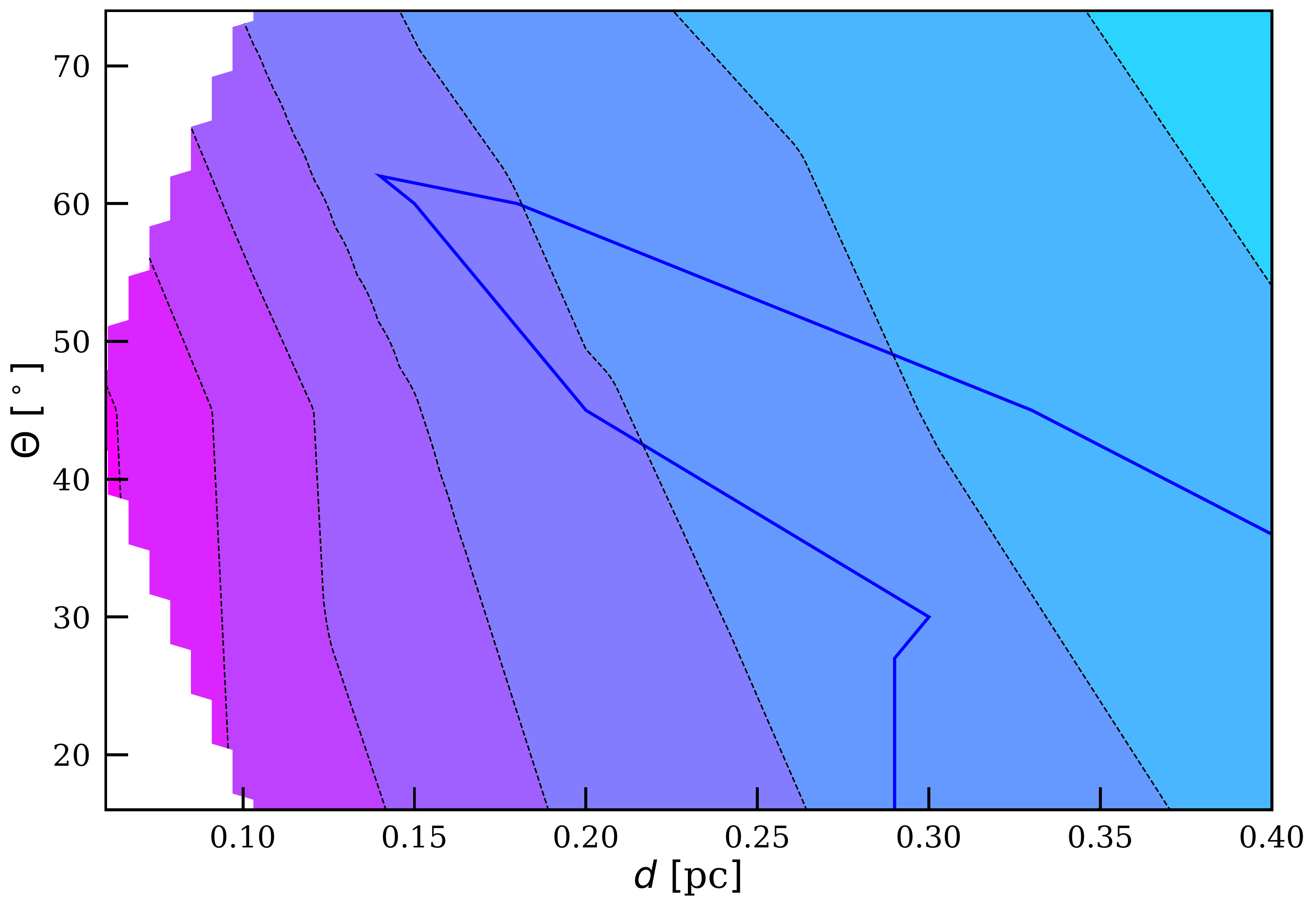}
\caption{Disk enrichment mass-fraction as function of the distance to
  the supernova and its angle with respect to the proto-planetary
  disk's rotation axis for SN11aof (left) and SN10a (right).
  The colored contours give the amount of accreted mass from the
  supernova remnant as fraction of the mass of the surviving disk
  (scale to the right). The enclosed and shaded regions are as in
  Fig.\,\ref{Fig:disk_heatedfraction}.  The small top-left area of the
  hashed region which overlaps with an enrichment of at least $m_{\rm
    shell}/M_{\rm disk} \apgt 10^{-3}$ satisfies all the criteria of
  the model.  the absence of a hashed region in the left panel
  indicates that there is no consistent solution across all
  requirements in our simulations for SN10a.
  \label{Fig:disk_enrichment}
}
\end{center}
\end{figure*}

A small amount of material from the nuclear blast wave is deposited in
the Sun's proto planetary disk.  We illustrate this in
Fig.\,\ref{Fig:disk_enrichment} for both supernova SN11aof and SN10a.  The
enclosed region indicates the area in distance to the supernova and
impact angle for which the truncation of the disk due to the passing
blast wave and the obliquity of the Sun with respect to the disk is
consistent with the observations.  The absence of a hashed region for
SN10a (right panel) indicates that in no part of the initial parameter
space for which the irradiation temperature of the disk due to the
supernova explosion reaches a temperature $\apgt 1500$\,K.

For a nearby supernovae $d \aplt 0.12$\,pc the amount of accreted
material can exceed $10^{-5}$ of the bound disk mass, but this
fraction drops rapidly with distance, which is consistent with the
earlier simulations by \cite{2007ApJ...662.1268O}.  A similar trend is
visible for SN11aof but the amount of accreted material is
considerably larger. For this latter supernova, the disk mass can be
enriched with $\apgt 10^{-6}$ up to a distance of $\sim 0.2$\,pc,
whereas for SN10a this amount is only achieved at a distance of $\aplt
0.12$\,pc. These results are consistent with earlier calculations of
disk enrichment by a nearby supernova
\cite{2007ApJ...662.1268O,2017MNRAS.469.1117C}.

\section{Discussion}\label{Sect:Discussion}

We simulated the supernova explosion and its effect on the
circumstellar disk in two stages.  In the first stage in
\S\,\ref{Sect:Irradiation} we perform the radiative transfer
calculation coupled to a hydrodynamical simulation in order to study
the effect of the irradiation of the supernova explosion on the Sun's
protoplanetary disk. In the second stage in \S\,\ref{Sect:blastwave},
we study the effect of the hot nuclear blast wave from the supernova
ejecta on the previously heated disk.

Our simulations are carried out with the AMUSE software framework,
which is designed for such multiphysics simulations. In our approach,
we coupled a radiative transfer solver with a hydrodynamics (SPH) code
and made them co-operate by interlaced temporal discretization.  We
performed extensive resolution and convergence tests to assure that
the individual codes produce satisfactory results, and we tested the
coupled codes for consistency.  It remains difficult, however, to
validate and verify such intricately coupled problems so long as there
is no analytic solution available.  Such complications hide in the
lack of a proper chemistry model, assumptions about coupling of grain
and gas temperatures, the choice of the underlying spectral energy
distribution of the adopted supernova model, and the state of the
protoplanetary disk (i.e.\, gaseous with some dust without ongoing
planet-formation).  Apart from the obvious theoretical extrapolations,
no such analytic solutions are readily available for the problem at
hand. Considering the stability of our results under small changes of
input physics and model parameters we at least argue that the results
presented here are robust and consistent.

The chains of events, first the heating of the disk by the supernova
radiation, followed by the blast wave and the consequent heating,
stripping and the injection of the supernova processed material may
have had important consequences for the formation and evolution of the
Solar System. We already argued that the inclination of the ecliptic
with respect to the Sun's equatorial plane and the truncation of the
current Kuiper belt could be the result of a nearby supernova.  Here
we further discuss some of the more speculative implications,
regarding the chemical evolution of the early Solar System.

\subsection{The heating of the disk}\label{Sect:heating}

In the range of distances and incident angles which satisfy the
observed size and tilt the disk is heated to about $1500$\,K (see the
hashed regions in Figs.\,\ref{Fig:temperatureSNPS11aof},
\ref{Fig:disk_heatedfraction} and Fig.\,\ref{Fig:disk_enrichment}).
After the heating, the disk rapidly cools to $\sim 100$\,K.  A few
decades later the supernova outflow material heats a considerable
fraction of the disk again to a comparable temperature.  In our model,
a fraction of the solids with up to $\sim 1$\,m size and the surfaces
of larger solids present in the Sun's circum-stellar disk would then
be heated to $\apgt 1500$\,K twice, once by radiation and again a few
decades later by the injection of the hot nuclear blast-wave. 

The second episode of heating is less global for the disk because it
casts a shadow on itself. In Fig.\,\ref{Fig:disk_heatedfraction} we
show that only a fraction of the disk is heated, which is the result
of the self-shadowing of the disk.  Supernova SN10a is less bright than
SN11aof and the total mass in the blast wave is smaller. Because the
out-flowing material has a higher velocity the fraction of the
disk-mass that is heated is comparable.

After the blast wave has passed, the temperature of the disk drops
again, resulting in a cooling rate of about 3-30\,K/hour.  The mean
temperature $\apgt 1500$\,K reached in the disk reported here for both
interactions (radiation and blast wave) are interesting because we can
speculate if they may be related to the still elusive physical process
responsible for the existence of chondrules
\citep{2008Sci...320.1617A}. These are the major components of most
meteorites and consist of free-floating mm-size droplets that were
liquefied at high temperatures. The origin of the possible
flash-heating events that liquefied the chondrules is unknown and we
cannot exclude that this may be related to local effects resulting
from the global disk-supernova interaction investigated here. Another
interesting feature of chondrules is their cooling rates of $0.5$ to
$1000$\,K/h \citep{2012Sci...338..651C??connelly16}, which is
considerably slower than expected for isolated particles. This may
imply that during the cooling phase they were embedded into hot gas,
which would be a natural consequence of being heated by the supernova
blast wave.

On the other hand, also the observations that chondrules exhibit a
wide range of ages and were probably being processed several times
requires an explanation. Our models would naturally account for two
heating events in short succession. However, it cannot be excluded
that in a relatively dense cluster more than one supernova exploded
nearby the proto-solar disk. More detailed studies are required to
ascertain if this supernova origin for the heating needed to explain
the existence of chondrules may be feasible.

\subsection{The effect of the presence of dust in the circumstellar disk on its temperature}\label{Sect:Dust}

\begin{figure}
\includegraphics[width=1.0\columnwidth]{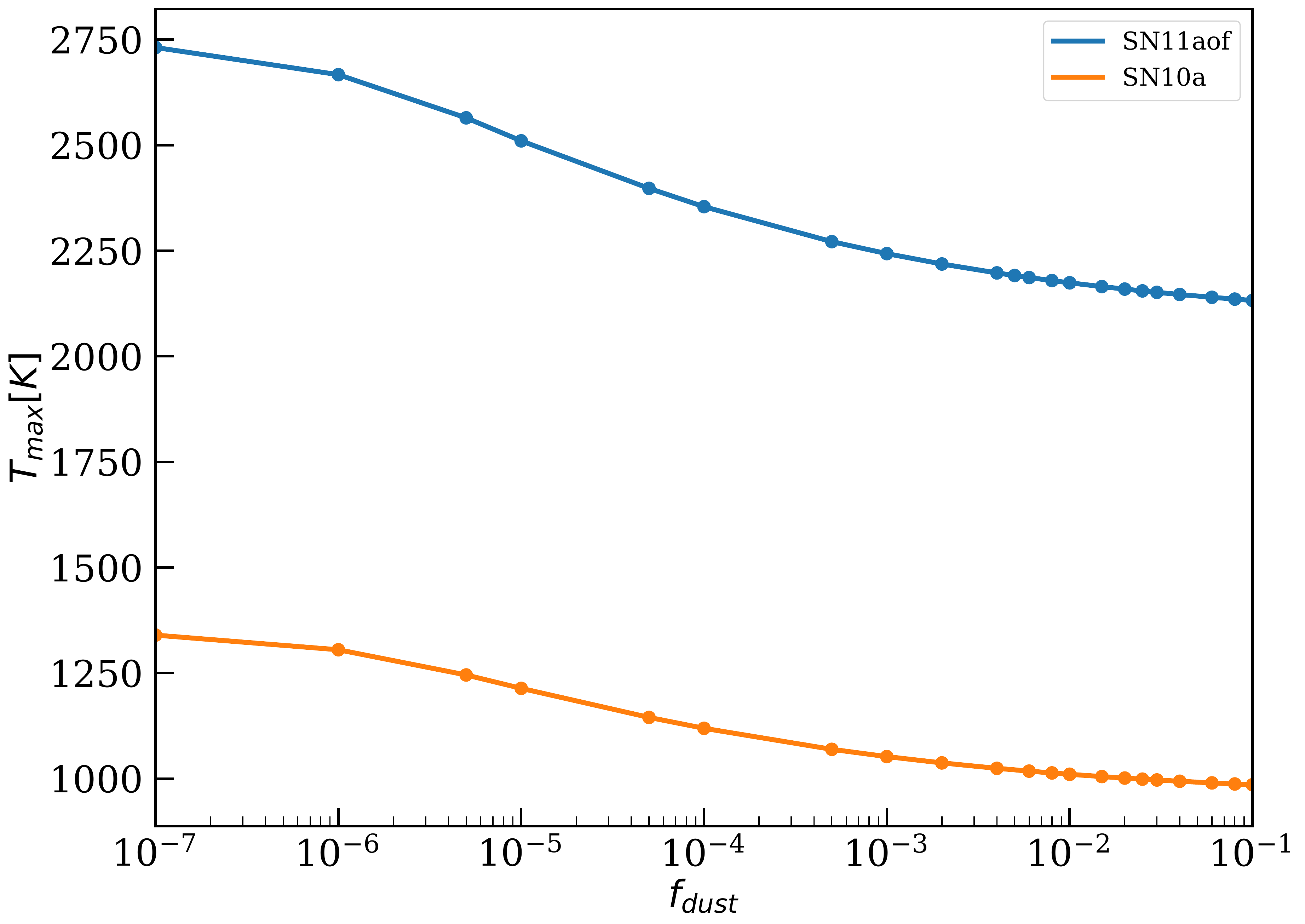}
\caption[]{Relative temperature as a function of the dust fraction.
    for inlincation $\Theta=0$ and at a distance of 0.2\,pc.
  \label{Fig:relative_dusttemp_vs_fdust}
}
\end{figure}
 
The calculations in \S\,\ref{Sect:RadiationTheDisk} ware performed
using the ray-tracing radiative transfer code {\tt SPHRay} without
support for dust extinction.  The presence of dust is expected to
reduce the temperature because it is efficient for storing and
subsequently re-radiating the heat. To test this hypothesis we perform
a new set of calculations in which we include the effect of dust in
the effective cross section per hydrogen atom
\begin{equation}
\sigma(\nu)  = \sigma_0 \xi(\nu) \left( f_{\rm dust} \over 0.01 \right).
\end{equation}
Here we adopt $\xi(\nu)=3$ is constant and $\sigma_0=5.0\times
10^{-22}$\,cm$^-2$.  In Fig.\,\ref{Fig:relative_dusttemp_vs_fdust} we
present the effect of the dust-to-gass mass fraction on the
temperature of the circumstellar disk at the peak of the supernova
luminosity (26 days after the supernova for SN11aof, and 10 days for
SN10a.  In these cases, we adopted $\Theta=0$ and a distance of
0.2\,pc.  The effect of dust is to generally reduce the mean
temperature of the disk at any moment, although we only present the
peak mean temperature in
Fig.\,\ref{Fig:relative_dusttemp_vs_fdust}. For a small dust fraction
of $f_{\rm dust} \aplt 10^{-7}$ the temperature saturates to the
no-dust case presented in \S\,\ref{Sect:RadiationTheDisk}, but for
higher dust fractions the temperature drops by up to about 20\%.

\begin{figure*}
\begin{center}
\includegraphics[width=1.0\columnwidth]{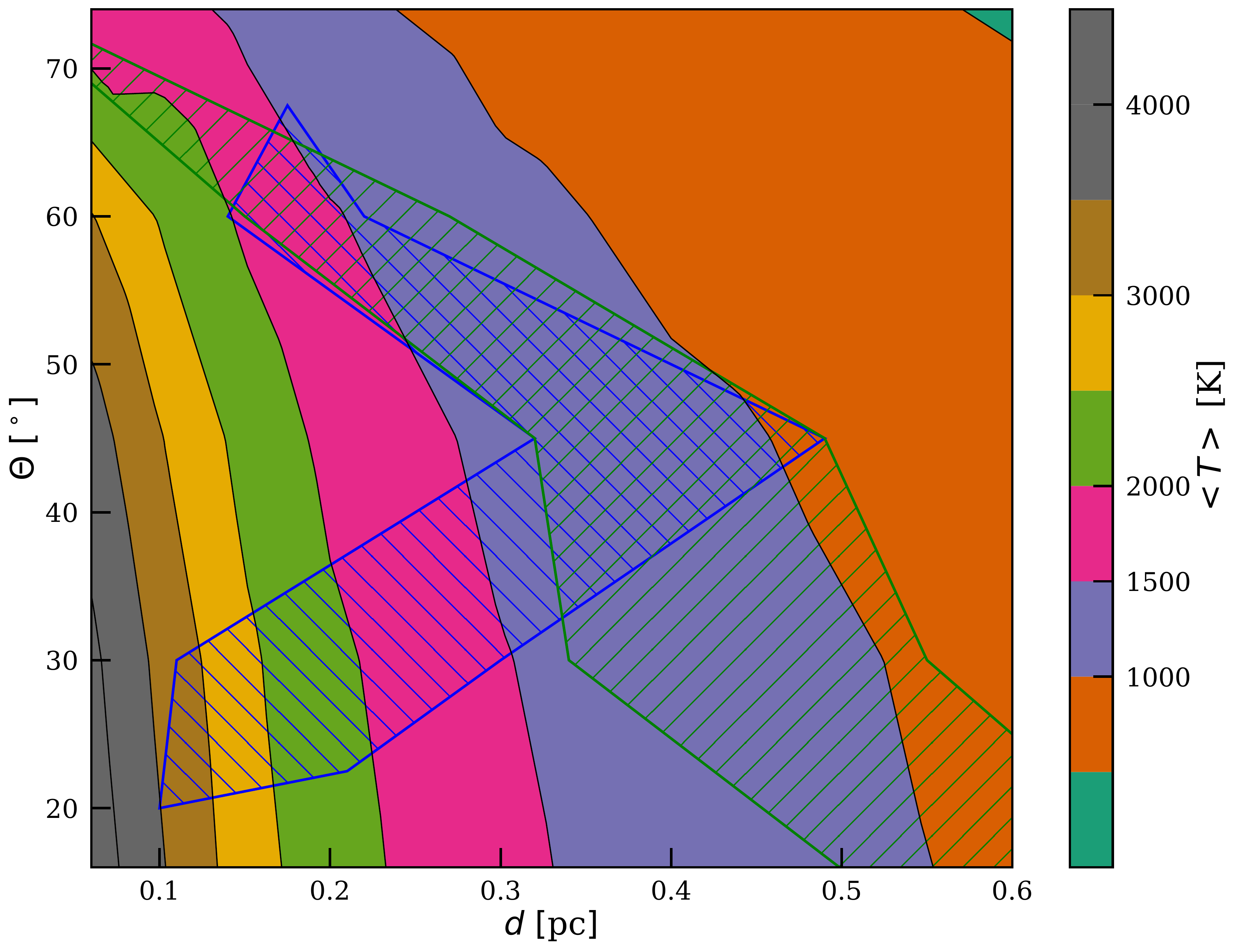}
~\includegraphics[width=1.0\columnwidth]{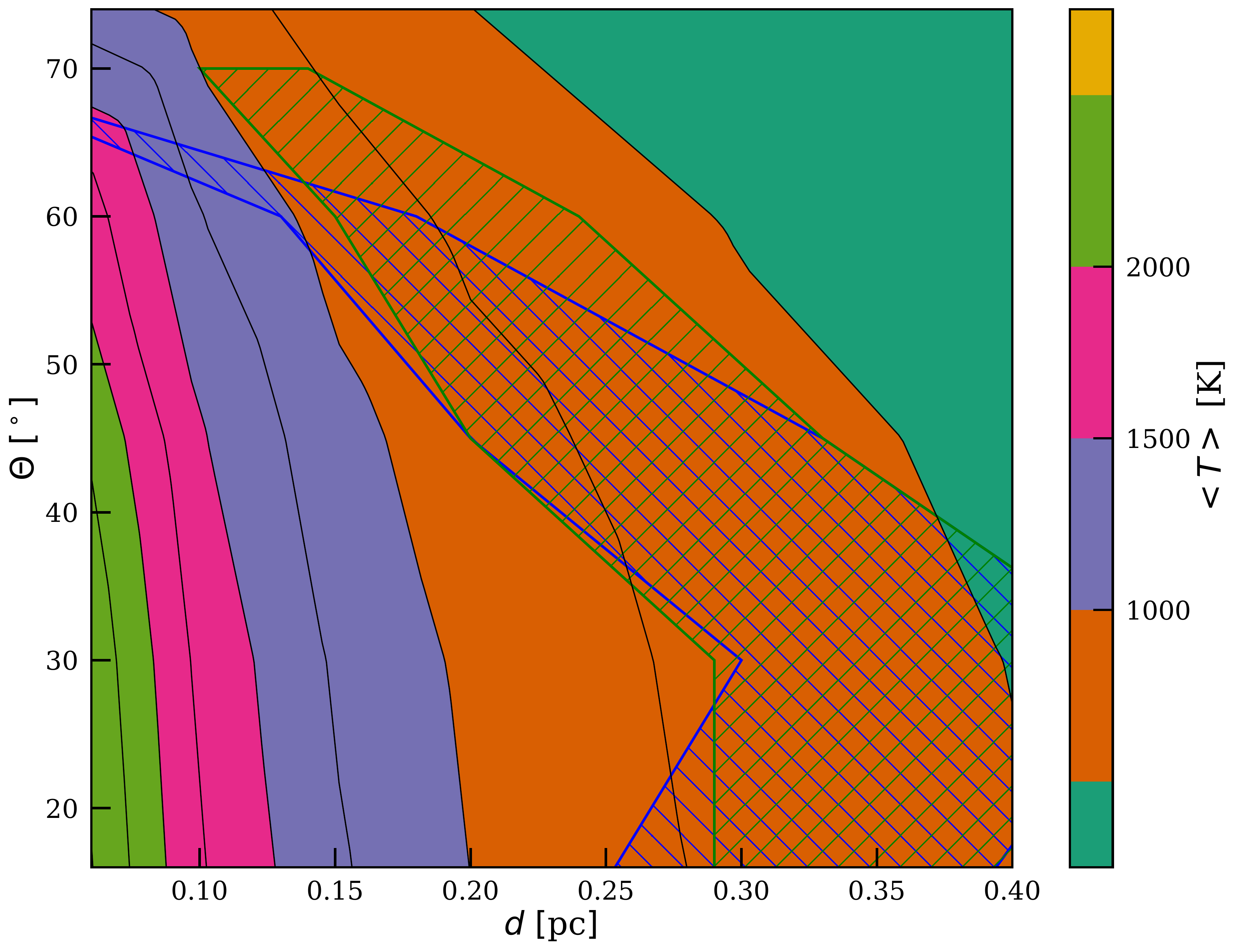}
\caption{Peak mean temperature of the disk (contours) due to the
  supernova radiation as a function of the distance to a supernova
  ($d$ in parsec) and the impact angle ($\Theta$ in degrees, measured
  from the disk's angular-momentum axis).  The adopted type IIp
  supernovae, SN11aof (left) has a peak luminosity of $1.1 \times
  10^{43}$\,erg/s and SN10a (right) has a peak luminosity of $2.3
  \times 10^{42}$\,erg/s.  The two shaded regions indicate the range
  in distance for which the Sun's obliquity is consistent with the
  currently observed value of $5^\circ.6\pm1^\circ.2$ (blue with
  ``//'' hashes), and for which the disk is truncated to 42--55\,\AU\,
  (green with hashes ).  This area is indicated in
  Fig.\,\ref{Fig:disk_enrichment} with a blue polygon.  The most
  promising part of parameter space is where the two hashed areas
  overlap because for there both criteria are met.
  \label{Fig:temperatureSNPS11aof_with_dust}
  \label{Fig:temperatureSNPS10a_with_dust}
}
\end{center}
\end{figure*}

We repeated the calculations in \S\,\ref{Sect:RadiationTheDisk} with a
constant gas-to-dust ratio $f_{\rm dust}=0.01$. The results of these
calculations are presented in
Fig.\,\ref{Fig:temperatureSNPS11aof_with_dust}.  The presence of dust
in the circumstellar disk results in an overall lower mean temperature
over the entire parameter space in $\Theta$ and $d$. These differences
may have a profound effect on the disk chemistry, but for our
discussion, the effect is not profound and does not affect our
conclusions.

\subsection{The accretion of short-lived radionuclids}

In order to explain the abundance of \Al\, observed in the early Solar
System a certain fraction of material from the supernova needs to be
incorporated in the disk. From the derived value of the
$^{26}$Al/$^{27}$Al ratio of $5 \times 10^{-5}$
\citep{2008LPI....39.1999J} and the recommended $^{27}$Al abundance of
the proto-sun of $6.26 \times 10^{-5}$ \citep[adopting a value of
  $A(E1)_0 = 6.54$,][]{lodders09} the early Solar system disk of
0.01\,\MSun\, is expected to be enriched by a relative mass-fraction
of $m_{\rm {26}_{Al}}/m_{\rm disk} \simeq 3 \times 10^{-9}$ of \Al.  A
25\,\MSun\, star has a \Al\, yield of $\sim 5 \times 10^{-5}$\,\MSun\,
\citep{TurHegerAustin2010,lugaro14}.  In order to be able to accrete
the observed amount of \Al\, by the supernova blast-wave a total mass
fraction of $m_{\rm accreted}/m_{\rm disk} \simeq 6 \times
10^{-5}$\,\MSun\, has to be accreted, which exceeds the fraction of
mass accreted in our simulations (see
Fig.\,\ref{Fig:disk_enrichment}).

Considering this value and the amount of material accreted by the
circumstellar disk (see Fig.\,\ref{Fig:disk_enrichment}) we conclude
that only for an extremely nearby supernova $\aplt 0.1$\,pc and at a
relatively small angle with respect to the disk $\aplt 40^\circ$ a
sufficient amount of \Al can be accreted. The disadvantage of such a
nearby supernova would be the almost complete obliteration of the
disk.  Our result is consistent with the earlier calculations of
\citet{2007ApJ...662.1268O,2017MNRAS.464.4318N} who argue that
injecting SLRs into a protosolar disk from a
nearby supernova is not straightforward and may require an early
condensation of \Al\, in large dust grains that can be accreted more
easily by the disk \citep{2010ApJ...711..597O}.

The abundance of $^{60}$Fe in the early Solar System may be too low to
allow for a nearby supernova, although such abundance is still debated
actively.  It is difficult to separate $^{27}$Al and $^{60}$Fe in
supernova ejecta because these two nuclei are produced in the same
regions \citep{timmes95,limongi06}. Furthermore, proof that both
nuclei are produced by massive stars is provided by $\gamma$-ray
spectroscopic observations \citep{diehl13}.  The flux ratio observed
$^{60}$Fe/$^{27}$Al of roughly 0.15 \citep{wang07} corresponds to an
abundance ratio of 0.55, which is a few to orders of magnitude higher
than the ratios measured for the early Solar System. This indicates
that the early Solar System did not sample the average Galactic
content of SLRs. In addition, evolutionary models
for massive stars have some difficulties to avoid overproducing even
the Galactic ratio \citep{limongi06,2016ApJ...821...38S,austin17}.

Following \citet{2010ApJ...711..597O}, we can still assume that the
required amount of \Al\, was incorporated into the early Solar System
because it was carried into the disk inside large dust grains. In this
case, if the highest observed values for \Fe\, are representative of
the early Solar System, then the scenario of disk injection from the
supernova is still plausible. Even in that case, however, it is not
clear to what extend the shock-wave of the interaction between the
supernova and the disk can be responsible for the origin of
chondrules.

To answer this question we consider time scales. Aluminum-26 is
observed to be present in the oldest components of meteorites, which
are called calcium-aluminum inclusions \citep{2008E&PSL.272..353J},
and are distributed within the entire disk \citep{2009Sci...325..985V}
with abundance ratios varying at least within a factor of two
\citep{2011ApJ...735L..37L}. Chondrules, on the other hand, are
expected to have formed over a time scale of a few Myr, beginning with
the moment the first CAIs formed \citep{2012Sci...338..651C}. This
implies that the fist heating event modeled here could not have
contributed to the production of chondrules since it occurred before
the \Al\, survival within the blast wave responsible for the second
heating event. The first heating event may still have been responsible
for the melting of some CAIs (only those which are \Al-poor) also show
igneous features, but for this, there seems to be no clear relation to
chondrules \citep[e.g.][]{2017GeCoA.196...18N}. If the interaction
with the blast wave was responsible for both the origin of SLRs and
the heating event that produced the oldest chondrules, then we are
left with the problem that future heating events are needed to produce
chondrules for the next few more Myr. If these were also supernova
related, then they should have been poor in \Al\, otherwise the
abundance variations in chondrules would have been considerably
larger.

It appears difficult to argue that the same supernova that is
responsible for the presence of \Al\, in the early Solar System was
also responsible for the shock waves responsible for the formation of
chondrules. As we have demonstrated here the interaction between a
supernova and a disk that could have injected SLRs in the early Solar
system would also have heated a considerable fraction of the disk to
temperatures sufficiently high to melt even the hardest dust
grains. This would have profound consequences for the further chemical
evolution of the disk. At this moment we do not know what these
consequences could entail, but this remains to be studied.

Another important issue to consider is that supernova SN11aof was
produced by a star of at least 25\,\MSun.  Such massive stars take
$\aplt 8$\,million years to evolve.  This is longer than the age
spread of $<3$\,Myr observed in the nearby star-forming regions
W3~\citep{2012ApJ...744...87B} and the Orion Nebular
Cluster~\citep{2011MNRAS.418.1948J}.  Although this would indicate a
rather prolonged disk lifetime, it is consistent with recent findings
regarding the average disk lifetime of planetary systems being longer
than 10\,Myr as observed in the large Magellanic cloud
\citep{2010ApJ...715....1D}, or the presence of a disk around the
$\sim 12$\,Myr old star TW\,Hydra \citep{2016ApJ...819L...7N}.  We,
therefore, argue that by the time a nearby star exploded the Sun and
its disk may still have had an age similar to that of the exploding
star, modulo the spread in the star-formation history.

\subsection{Inplications for other planetary systems}

For the planetary systems that are born in clusters of $500$ to $10^4$
stars the interaction between a supernova blast-wave and the
circumstellar disk may be rather common. Although we only cover a
rather limited parameter space in terms of supernova type, distance
and incident angle, we compare the distribution of observed
inclinations with the simulations.

We used the last 10 snapshots for each simulation in the grid of
parameter space ($d$ and $\Theta$) to construct a cumulative
distribution of the measured inclination. We weighted the simulations
based on the distance and inclination.  For the distance, we adopted a
\cite{1911MNRAS..71..460P} distribution with a characteristic length
of 1.0\,pc. The underlying assumption is that the star cluster, at an
age of 10\,Myr, resembles a Plummer distribution with the exploding
star in the cluster center.  The inclinations were weighted with
$\cos(\Theta)$.

\begin{figure}
\begin{center}
\includegraphics[width=1.0\columnwidth]{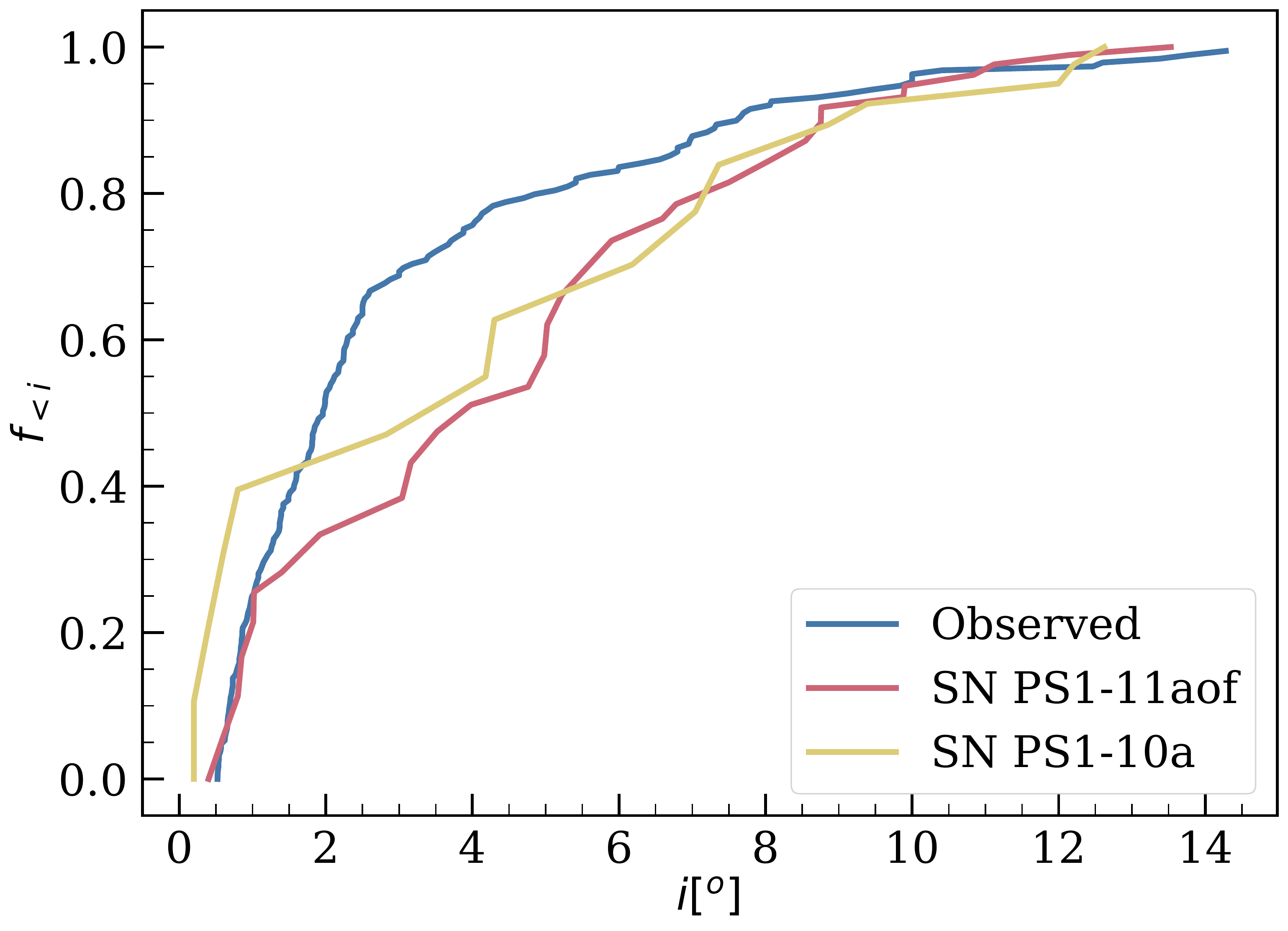}
\end{center}
\caption{Cumulative distribution of the inclination for observed
  planetary systems (blue) with our two supernova models SN11aof
  (red) and SN10a (yellow). The two simulated distributions are
  statistically indistinguishable (p-value $= 0.47$), but both
  distributions compare less favorable with the observations (p-value
  $\sim 0.011$).
  \label{Fig:inclination_distribution}
}
\end{figure}

For the observed systems, we adopt the Kepler database\footnote{see
  {\tt https://exoplanetarchive.ipac.caltech.edu/}} from July 2017.
We selected the systems for which the rotation of the parent star and
the inclination of at least one planet was observed.  In a subsequent
pass we selected the 189 systems for which this inclination angle was
$1^\circ < i < 15^\circ$.  Their cumulative distribution is presented
in Fig.\,\ref{Fig:inclination_distribution} (blue curve).  The
simulated distributions (red and yellow curves in
Fig.\,\ref{Fig:inclination_distribution}) are statistically
indistinguishable but differ from the observed distribution (with a
Kolmogorov-Smirnoff \citep{1954Kolmogorov} p-value of 1.1\%).  The
Kepler systems to which we compare, however, are all rather tight
compared to our 100\,\AU\, sized disks and in addition, the
interaction with a supernova blast-wave is probably not the only
mechanism that can induce a tilt in a circumstellar disk.  Other
mechanisms that may be important in affecting the obliquity of the
star with respect to its planetary disk include close stellar
encounters and secular dynamics, both of which are probably important
in the same environment \citep{2017MNRAS.470.4337C}.

\section{Conclusions}

We performed radiative transfer and hydrodynamical simulations of the
interaction between a supernova and a Sun-like star with a
circumstellar disk. The calculations are performed using the AMUSE
software environment and all the scripts used in this manuscript are
available at the package website at \url{amusecode.org} and at 
\url{github.com/amusecode}.

We have discussed the relevance of our model to cosmochemistry
constraints, i.e., to the presence of SLRs in the
early Solar System and to the shock heating events possibly
responsible for the origin of chondrules.  We confirm, as argued by
\cite{2007ApJ...662.1268O}, that the amount of SLRs with which a protoplanetary disk would be enriched is too
small to explain their abundance in the Solar System.  They argue that
the incorporation of \Al\, and \Fe\, in large dust grains, would allow
these elements to stick to the disk, but they do not explain why \Fe\,
would be underabundance in this process.  More detailed models of the
local impact of the heat generated by the supernova-disk interaction
are neede to assess if this may play a role in the origin of
chondrules.

We identified two distinct physical consequences that provide
independent constraints on the distance between the exploding star
and the Solar System, and on the relative inclination-angle of the
supernova with respect to the disk's angular momentum axis. The Solar
System's circum-stellar disk is sensitive to these parameters via the
ablation of the circumstellar disk and the tilt induced by the blast
wave of the supernova shell.  By comparing the observed disk size and
obliquity of the Sun with respect to the ecliptic with the simulations
we then constrain the type of supernova, its distance and the angle
with respect to the Sun's circumstellar disk.  Consistent results
between observations and simulations are obtained when the supernova
occurred at a distance of $0.15 \aplt d \aplt 0.40$\,pc, and at an
angle of $35^\circ \aplt \Theta \aplt 65^\circ$. We only tried two
supernovae, both are of type IIp, and it is hard to be conclusive, but
we prefer supernova SN11aof because it is brighter and it has a
higher-density mass-outflow, but at a somewhat lower velocity than the
other supernova we tried, SN10a.  We also confirm that the amount of
blast-wave material that is accreted by the circumstellar disk is
insufficient to explain the enrichment of short-lived radionuclides.

Considering all constraints, we tend to prefer the brighter supernova
SN11aof to explain the temperature evolution of the protoplanetary disk,
the Sun's obliquity and the outer disk radius. This supernova would
have occurred at a distance of 0.15--0.25\,pc and from an angle of
$52^\circ$ to $66^\circ$.  However, if the temperature cannot be used
to constrain the initial conditions we have to relax the viable
initial parameter space to $d \simeq 0.15$ to $0.40$\,pc and $\Theta
\simeq 35^\circ$ to about $66^\circ$.

{\bf Acknowledgements} It is a pleasure to thank Ignas Snellen,
Vincent Icke, Mher Kazandjian, and Xander Tielens, Guido de Marchi,
Marco Delbo, Shu-Ichiro Inutsuka and Eiichiro Kokubo.  This work was
supported by the Netherlands Research School for Astronomy (NOVA), NWO
(grant \#621.016.701 [LGM-II]) and by the European Research Council
(ERC) under the EU Horizon 2020 research and innovation programme
(grant agreement No 724560) through ERC Consolidator Grant
``RADIOSTAR'' to M.L.


\begin{thebibliography}{65}
\expandafter\ifx\csname natexlab\endcsname\relax\def\natexlab#1{#1}\fi

\bibitem[{{Alexander} {et~al.}(2008){Alexander}, {Grossman}, {Ebel}, \&
  {Ciesla}}]{2008Sci...320.1617A}
{Alexander}, C.~M.~O.~., {Grossman}, J.~N., {Ebel}, D.~S., \& {Ciesla}, F.~J.
  2008, Science, 320, 1617

\bibitem[{{Altay} {et~al.}(2008){Altay}, {Croft}, \&
  {Pelupessy}}]{2008MNRAS.386.1931A}
{Altay}, G., {Croft}, R.~A.~C., \& {Pelupessy}, I. 2008, \mnras, 386, 1931

\bibitem[{{Altay} {et~al.}(2011){Altay}, {Croft}, \&
  {Pelupessy}}]{2011ascl.soft03009A}
{Altay}, G., {Croft}, R.~A.~C., \& {Pelupessy}, I. 2011, {SPHRAY: A Smoothed
  Particle Hydrodynamics Ray Tracer for Radiative Transfer}, Astrophysics
  Source Code Library

\bibitem[{{Andrews} {et~al.}(2009){Andrews}, {Wilner}, {Hughes}, {Qi}, \&
  {Dullemond}}]{2009ApJ...700.1502A}
{Andrews}, S.~M., {Wilner}, D.~J., {Hughes}, A.~M., {Qi}, C., \& {Dullemond},
  C.~P. 2009, \apj, 700, 1502

\bibitem[{{Andrews} {et~al.}(2010){Andrews}, {Wilner}, {Hughes}, {Qi}, \&
  {Dullemond}}]{2010ApJ...723.1241A}
{Andrews}, S.~M., {Wilner}, D.~J., {Hughes}, A.~M., {Qi}, C., \& {Dullemond},
  C.~P. 2010, \apj, 723, 1241

\bibitem[{{Armitage}(2011)}]{2011ARA&A..49..195A}
{Armitage}, P.~J. 2011, \araa, 49, 195

\bibitem[{{Austin} {et~al.}(2017){Austin}, {West}, \& {Heger}}]{austin17}
{Austin}, S.~M., {West}, C., \& {Heger}, A. 2017, \apj, 839, L9

\bibitem[{{Beck} \& {Giles}(2005)}]{2005ApJ...621L.153B}
{Beck}, J.~G. \& {Giles}, P. 2005, \apjl, 621, L153

\bibitem[{{Bik} {et~al.}(2012){Bik}, {Henning}, {Stolte}, {Brandner},
  {Gouliermis}, {Gennaro}, {Pasquali}, {Rochau}, {Beuther}, {Ageorges},
  {Seifert}, {Wang}, \& {Kudryavtseva}}]{2012ApJ...744...87B}
{Bik}, A., {Henning}, T., {Stolte}, A., {et~al.} 2012, \apj, 744, 87

\bibitem[{{Cai} {et~al.}(2017){Cai}, {Kouwenhoven}, {Portegies Zwart}, \&
  {Spurzem}}]{2017MNRAS.470.4337C}
{Cai}, M.~X., {Kouwenhoven}, M.~B.~N., {Portegies Zwart}, S.~F., \& {Spurzem},
  R. 2017, \mnras, 470, 4337

\bibitem[{{Close} \& {Pittard}(2017)}]{2017MNRAS.469.1117C}
{Close}, J.~L. \& {Pittard}, J.~M. 2017, \mnras, 469, 1117

\bibitem[{{Connelly} {et~al.}(2012){Connelly}, {Bizzarro}, {Krot}, {Nordlund},
  {Wielandt}, \& {Ivanova}}]{2012Sci...338..651C}
{Connelly}, J.~N., {Bizzarro}, M., {Krot}, A.~N., {et~al.} 2012, Science, 338,
  651

\bibitem[{{Davis} \& {Richter}(2005)}]{2005mcp..book..407D}
{Davis}, A.~M. \& {Richter}, F.~M. 2005, {Condensation and Evaporation of Solar
  System Materials}, ed. A.~M. {Davis}, H.~D. {Holland}, \& K.~K. {Turekian}
  (Elsevier B), 407

\bibitem[{{De Marchi} {et~al.}(2010){De Marchi}, {Panagia}, \&
  {Romaniello}}]{2010ApJ...715....1D}
{De Marchi}, G., {Panagia}, N., \& {Romaniello}, M. 2010, \apj, 715, 1

\bibitem[{{Diehl}(2013)}]{diehl13}
{Diehl}, R. 2013, Reports on Progress in Physics, 76, 026301

\bibitem[{{Dwarkadas} {et~al.}(2017){Dwarkadas}, {Dauphas}, {Meyer},
  {Boyajian}, \& {Bojazi}}]{dwarkadas17}
{Dwarkadas}, V.~V., {Dauphas}, N., {Meyer}, B., {Boyajian}, P., \& {Bojazi}, M.
  2017, \apj, 851, 147

\bibitem[{{Faran} {et~al.}(2018){Faran}, {Nakar}, \&
  {Poznanski}}]{2018MNRAS.473..513F}
{Faran}, T., {Nakar}, E., \& {Poznanski}, D. 2018, \mnras, 473, 513

\bibitem[{{Gaidos} {et~al.}(2009){Gaidos}, {Krot}, \&
  {Huss}}]{2009ApJ...705L.163G}
{Gaidos}, E., {Krot}, A.~N., \& {Huss}, G.~R. 2009, \apjl, 705, L163

\bibitem[{{Gerritsen} \& {Icke}(1997)}]{1997A&A...325..972G}
{Gerritsen}, J.~P.~E. \& {Icke}, V. 1997, \aap, 325, 972

\bibitem[{{Gounelle} \& {Meynet}(2012)}]{gounelle12}
{Gounelle}, M. \& {Meynet}, G. 2012, \aap, 545, A4

\bibitem[{{Hernquist} \& {Katz}(1989)}]{1989ApJS...70..419H}
{Hernquist}, L. \& {Katz}, N. 1989, \apjs, 70, 419

\bibitem[{{Iliev} {et~al.}(2006){Iliev}, {Ciardi}, {Alvarez}, {Maselli},
  {Ferrara}, {Gnedin}, {Mellema}, {Nakamoto}, {Norman}, {Razoumov},
  {Rijkhorst}, {Ritzerveld}, {Shapiro}, {Susa}, {Umemura}, \&
  {Whalen}}]{2006MNRAS.371.1057I}
{Iliev}, I.~T., {Ciardi}, B., {Alvarez}, M.~A., {et~al.} 2006, \mnras, 371,
  1057

\bibitem[{{Jacobsen} {et~al.}(2008{\natexlab{a}}){Jacobsen}, {Yin}, {Moynier},
  {Amelin}, {Krot}, {Nagashima}, {Hutcheon}, \& {Palme}}]{2008E&PSL.272..353J}
{Jacobsen}, B., {Yin}, Q.-z., {Moynier}, F., {et~al.} 2008{\natexlab{a}}, Earth
  and Planetary Science Letters, 272, 353

\bibitem[{{Jacobsen} {et~al.}(2008{\natexlab{b}}){Jacobsen}, {Chakrabarti},
  {Ranen}, \& {Petaev}}]{2008LPI....39.1999J}
{Jacobsen}, S.~B., {Chakrabarti}, R., {Ranen}, M.~C., \& {Petaev}, M.~I.
  2008{\natexlab{b}}, in Lunar and Planetary Science Conference, Vol.~39, Lunar
  and Planetary Science Conference, 1999

\bibitem[{{Jeffries} {et~al.}(2011){Jeffries}, {Littlefair}, {Naylor}, \&
  {Mayne}}]{2011MNRAS.418.1948J}
{Jeffries}, R.~D., {Littlefair}, S.~P., {Naylor}, T., \& {Mayne}, N.~J. 2011,
  \mnras, 418, 1948

\bibitem[{{Kolmogorov}(1954)}]{1954Kolmogorov}
{Kolmogorov}, A.~N. 1954, Dolk Akad.\, Nauk SSSR, 98, 527

\bibitem[{{Lada} \& {Lada}(2003)}]{2003ARA&A..41...57L}
{Lada}, C.~J. \& {Lada}, E.~A. 2003, \araa, 41, 57

\bibitem[{{Larsen} {et~al.}(2011){Larsen}, {Trinquier}, {Paton}, {Schiller},
  {Wielandt}, {Ivanova}, {Connelly}, {Nordlund}, {Krot}, \&
  {Bizzarro}}]{2011ApJ...735L..37L}
{Larsen}, K.~K., {Trinquier}, A., {Paton}, C., {et~al.} 2011, \apjl, 735, L37

\bibitem[{{Lichtenberg} {et~al.}(2016){Lichtenberg}, {Parker}, \&
  {Meyer}}]{2016MNRAS.462.3979L}
{Lichtenberg}, T., {Parker}, R.~J., \& {Meyer}, M.~R. 2016, \mnras, 462, 3979

\bibitem[{{Limongi} \& {Chieffi}(2006)}]{limongi06}
{Limongi}, M. \& {Chieffi}, A. 2006, \apj, 647, 483

\bibitem[{{Lodders} {et~al.}(2009){Lodders}, {Palme}, \& {Gail}}]{lodders09}
{Lodders}, K., {Palme}, H., \& {Gail}, H.-P. 2009, Landolt-B{\"o}rnstein, New
  Series VI/4B, 34, Chapter 4.4. [\eprint[arXiv]{0901.1149}]

\bibitem[{{Looney} {et~al.}(2006){Looney}, {Tobin}, \&
  {Fields}}]{2006ApJ...652.1755L}
{Looney}, L.~W., {Tobin}, J.~J., \& {Fields}, B.~D. 2006, \apj, 652, 1755

\bibitem[{{Lugaro} {et~al.}(2014){Lugaro}, {Heger}, {Osrin}, {Goriely},
  {Zuber}, {Karakas}, {Gibson}, {Doherty}, {Lattanzio}, \& {Ott}}]{lugaro14}
{Lugaro}, M., {Heger}, A., {Osrin}, D., {et~al.} 2014, Science, 345, 650

\bibitem[{{Meyer}(2005)}]{2005ASPC..341..515M}
{Meyer}, B.~S. 2005, in Astronomical Society of the Pacific Conference Series,
  Vol. 341, Chondrites and the Protoplanetary Disk, ed. A.~N. {Krot}, E.~R.~D.
  {Scott}, \& B.~{Reipurth}, 515

\bibitem[{{Mishra} \& {Chaussidon}(2014)}]{mishra14}
{Mishra}, R.~K. \& {Chaussidon}, M. 2014, Earth and Planetary Science Letters,
  398, 90

\bibitem[{{Needham} {et~al.}(2017){Needham}, {Messenger}, {Han}, \&
  {Keller}}]{2017GeCoA.196...18N}
{Needham}, A.~W., {Messenger}, S., {Han}, J., \& {Keller}, L.~P. 2017, \gca,
  196, 18

\bibitem[{{Nicholson} \& {Parker}(2017)}]{2017MNRAS.464.4318N}
{Nicholson}, R.~B. \& {Parker}, R.~J. 2017, \mnras, 464, 4318

\bibitem[{{Nomura} {et~al.}(2016){Nomura}, {Tsukagoshi}, {Kawabe}, {Ishimoto},
  {Okuzumi}, {Muto}, {Kanagawa}, {Ida}, {Walsh}, {Millar}, \&
  {Bai}}]{2016ApJ...819L...7N}
{Nomura}, H., {Tsukagoshi}, T., {Kawabe}, R., {et~al.} 2016, \apjl, 819, L7

\bibitem[{{Ouellette} {et~al.}(2007){Ouellette}, {Desch}, \&
  {Hester}}]{2007ApJ...662.1268O}
{Ouellette}, N., {Desch}, S.~J., \& {Hester}, J.~J. 2007, \apj, 662, 1268

\bibitem[{{Ouellette} {et~al.}(2010){Ouellette}, {Desch}, \&
  {Hester}}]{2010ApJ...711..597O}
{Ouellette}, N., {Desch}, S.~J., \& {Hester}, J.~J. 2010, \apj, 711, 597

\bibitem[{{Pelupessy} {et~al.}(2004){Pelupessy}, {van der Werf}, \&
  {Icke}}]{2004A&A...422...55P}
{Pelupessy}, F.~I., {van der Werf}, P.~P., \& {Icke}, V. 2004, \aap, 422, 55

\bibitem[{{Pelupessy} {et~al.}(2013){Pelupessy}, {van Elteren}, {de Vries},
  {McMillan}, {Drost}, \& {Portegies Zwart}}]{2013A&A...557A..84P}
{Pelupessy}, F.~I., {van Elteren}, A., {de Vries}, N., {et~al.} 2013, \aap,
  557, A84

\bibitem[{{Plummer}(1911)}]{1911MNRAS..71..460P}
{Plummer}, H.~C. 1911, \mnras, 71, 460

\bibitem[{{Portegies Zwart}(2011)}]{2011ascl.soft07007P}
{Portegies Zwart}, S. 2011, {AMUSE: Astrophysical Multipurpose Software
  Environment}, Astrophysics Source Code Library

\bibitem[{{Portegies Zwart} \& {McMillan}(2017)}]{AMUSE}
{Portegies Zwart}, S. \& {McMillan}, S. 2017, {Astrophysical Recipes: the Art
  of AMUSE} (AAS IOP Astronomy)

\bibitem[{{Portegies Zwart} {et~al.}(2013){Portegies Zwart}, {McMillan}, {van
  Elteren}, {Pelupessy}, \& {de Vries}}]{2013CoPhC.183..456P}
{Portegies Zwart}, S., {McMillan}, S.~L.~W., {van Elteren}, E., {Pelupessy},
  I., \& {de Vries}, N. 2013, Computer Physics Communications, 183, 456

\bibitem[{{Portegies Zwart}(2009)}]{2009ApJ...696L..13P}
{Portegies Zwart}, S.~F. 2009, \apjl, 696, L13

\bibitem[{{Ritzerveld} \& {Icke}(2006)}]{2006PhRvE..74b6704R}
{Ritzerveld}, J. \& {Icke}, V. 2006, Phys.\, Rev.\, E, 74, 026704

\bibitem[{{Safronov}(1960)}]{1960AnAp...23..979S}
{Safronov}, V.~S. 1960, Annales d'Astrophysique, 23, 979

\bibitem[{{Sanders} {et~al.}(2015){Sanders}, {Soderberg}, {Gezari},
  {Betancourt}, {Chornock}, {Berger}, {Foley}, {Challis}, {Drout}, {Kirshner},
  {Lunnan}, {Marion}, {Margutti}, {McKinnon}, {Milisavljevic}, {Narayan},
  {Rest}, {Kankare}, {Mattila}, {Smartt}, {Huber}, {Burgett}, {Draper},
  {Hodapp}, {Kaiser}, {Kudritzki}, {Magnier}, {Metcalfe}, {Morgan}, {Price},
  {Tonry}, {Wainscoat}, \& {Waters}}]{2015ApJ...799..208S}
{Sanders}, N.~E., {Soderberg}, A.~M., {Gezari}, S., {et~al.} 2015, \apj, 799,
  208

\bibitem[{{Shakura} \& {Sunyaev}(1973)}]{1973A&A....24..337S}
{Shakura}, N.~I. \& {Sunyaev}, R.~A. 1973, \aap, 24, 337

\bibitem[{{Spicer} {et~al.}(1990){Spicer}, {Maran}, \&
  {Clark}}]{1990ApJ...356..549S}
{Spicer}, D.~S., {Maran}, S.~P., \& {Clark}, R.~W. 1990, \apj, 356, 549

\bibitem[{{Sukhbold} {et~al.}(2016){Sukhbold}, {Ertl}, {Woosley}, {Brown}, \&
  {Janka}}]{2016ApJ...821...38S}
{Sukhbold}, T., {Ertl}, T., {Woosley}, S.~E., {Brown}, J.~M., \& {Janka}, H.-T.
  2016, \apj, 821, 38

\bibitem[{{Tang} \& {Dauphas}(2012)}]{2012LPI....43.1703T}
{Tang}, H. \& {Dauphas}, N. 2012, in Lunar and Planetary Science Conference,
  Vol.~43, Lunar and Planetary Science Conference, 1703

\bibitem[{{Tang} \& {Dauphas}(2015)}]{tang15}
{Tang}, H. \& {Dauphas}, N. 2015, \apj, 802, 22

\bibitem[{{Telus} {et~al.}(2018){Telus}, {Huss}, {Nagashima}, {Ogliore}, \&
  {Tachibana}}]{telus18}
{Telus}, M., {Huss}, G.~R., {Nagashima}, K., {Ogliore}, R.~C., \& {Tachibana},
  S. 2018, \gca, 221, 342

\bibitem[{{Timmes} {et~al.}(1995){Timmes}, {Woosley}, {Hartmann}, {Hoffman},
  {Weaver}, \& {Matteucci}}]{timmes95}
{Timmes}, F.~X., {Woosley}, S.~E., {Hartmann}, D.~H., {et~al.} 1995, \apj, 449,
  204

\bibitem[{{Toomre}(1964)}]{1964ApJ...139.1217T}
{Toomre}, A. 1964, \apj, 139, 1217

\bibitem[{Tur {et~al.}(2010)Tur, Heger, \& A.M.}]{TurHegerAustin2010}
Tur, C., Heger, A., \& A.M., S. 2010, Production of 26Al, 44Ti and 60Fe in
  Supernovae-sensitivity to the helium burning rates, Event11th Symposium on
  Nuclei in the Cosmos, NIC 2010 - Heidelberg, Germany

\bibitem[{{Vicente} \& {Alves}(2005)}]{2005A&A...441..195V}
{Vicente}, S.~M. \& {Alves}, J. 2005, \aap, 441, 195

\bibitem[{{Villeneuve} {et~al.}(2009){Villeneuve}, {Chaussidon}, \&
  {Libourel}}]{2009Sci...325..985V}
{Villeneuve}, J., {Chaussidon}, M., \& {Libourel}, G. 2009, Science, 325, 985

\bibitem[{{Wang} {et~al.}(2007){Wang}, {Harris}, {Diehl}, {Halloin}, {Cordier},
  {Strong}, {Kretschmer}, {Kn{\"o}dlseder}, {Jean}, {Lichti}, {Roques},
  {Schanne}, {von Kienlin}, {Weidenspointner}, \& {Wunderer}}]{wang07}
{Wang}, W., {Harris}, M.~J., {Diehl}, R., {et~al.} 2007, \aap, 469, 1005

\bibitem[{{Wijnen} {et~al.}(2017){Wijnen}, {Pelupessy}, {Pols}, \& {Portegies
  Zwart}}]{2017A&A...604A..88W}
{Wijnen}, T.~P.~G., {Pelupessy}, F.~I., {Pols}, O.~R., \& {Portegies Zwart}, S.
  2017, \aap, 604, A88

\bibitem[{{Wijnen} {et~al.}(2016){Wijnen}, {Pols}, {Pelupessy}, \& {Portegies
  Zwart}}]{2016A&A...594A..30W}
{Wijnen}, T.~P.~G., {Pols}, O.~R., {Pelupessy}, F.~I., \& {Portegies Zwart}, S.
  2016, \aap, 594, A30

\bibitem[{{Wisdom} \& {Holman}(1991)}]{1991AJ....102.1528W}
{Wisdom}, J. \& {Holman}, M. 1991, \aj, 102, 1528

\end{thebibliography}
\end{document}